\documentclass[aps,prc,nofootinbib,groupedaddress,twocolumn,10pt,floatfix]{revtex4-2}

\usepackage{emnorm}

\ifproofpre{}{\ifeof18
    \typeout{Version control information not updated!}
\else
    \immediate\write18{vc/vc}
\fi

\RequirePackage{xcolor}

\IfFileExists{vc/vc-info.tex}{\gdef\GITAbrHash{6f3c668}\gdef\VCRevision{\GITAbrHash}\gdef\VCRevisionMod{\VCRevision}\gdef\VCRevisionMod{\VCRevision} }{
\gdef\GITAbrHash{unknown}\gdef\VCRevision{\GITAbrHash}\gdef\VCRevisionMod{\VCRevision}}
 }

\begin{document}

\title{Robust \textit{ab initio} prediction of nuclear electric quadrupole
  observables by scaling to the charge radius}

\author{Mark A.~Caprio}
\affiliation{Department of Physics and Astronomy, University of Notre Dame, Notre Dame, Indiana 46556-5670, USA}

\author{Patrick J.~Fasano}
\affiliation{Department of Physics and Astronomy, University of Notre Dame, Notre Dame, Indiana 46556-5670, USA}

\author{Pieter Maris}
\affiliation{Department of Physics and Astronomy, Iowa State University, Ames, Iowa 50011-3160, USA}

\date{June 18, 2022}

\begin{abstract}
  Meaningful predictions for electric quadrupole ($E2$) observables
from \textit{ab initio} nuclear theory are necessary, if the \textit{ab initio}
description of collective correlations is to be confronted with experiment, as
well as to provide predictive power for unknown $E2$ observables.
However, converged results for $E2$ observables are notoriously challenging
to obtain in \textit{ab initio} no-core configuration interaction (NCCI)
approaches.  Matrix elements of the $E2$ operator are sensitive to the
large-distance tails of the nuclear wave function, which converge slowly in an
oscillator basis expansion.
Similar convergence challenges beset \textit{ab initio} prediction of the
nuclear charge radius.  We demonstrate that the convergence patterns of the $E2$
and radius observables are strongly correlated, and that meaningful predictions
for the absolute scale of $E2$ observables may be made by calibrating to the
experimentally-known ground-state charge radius.
We illustrate by providing robust \textit{ab initio} predictions for several
$E2$ transition strengths and quadrupole moments in $p$-shell nuclei, in cases where
experimental results are available for comparison.
\relax
\relax
 \end{abstract}

\preprint{Git hash: \VCRevisionMod}

\maketitle

\paragraph*{Introduction.}
Electric quadrupole ($E2$) observables, including $E2$ transition strengths and
electric quadrupole moments, probe nuclear deformation and collective
structure~\cite{bohr1998:v2,casten2000:ns,rowe2010:collective-motion}.  The
absolute scale of $E2$ observables provides a measure of the overall
deformation~\cite{raman2001:systematics}, while their relative strengths
(\textit{e.g.}, for transitions among rotational
states~\cite{alaga1955:branching}) are indicative of the structural
characteristics of the excitation spectrum.

In \textit{ab initio} descriptions of light nuclei, a fully microscopic
description of the nuclear many-body problem is attempted directly in terms of
the nucleons and their free-space interactions.  Signatures of collective
phenomena, including
clustering~\cite{pieper2004:gfmc-a6-8,neff2004:cluster-fmd,maris2012:mfdn-ccp11,yoshida2013:ncmcsm-8be-10be-6be-cluster,romeroredondo2016:6he-correlations,navratil2016:ncsmc}
and
rotation~\cite{caprio2013:berotor,*maris2015:berotor2,*maris2019:berotor2-ERRATUM,caprio2015:berotor-ijmpe,stroberg2016:ab-initio-sd-multireference,jansen2016:sd-shell-ab-initio,caprio2020:bebands},
arise in the results.  To confront
these descriptions with experiment, it is necessary to obtain
concrete, quantitative predictions for $E2$ observables.  These serve both to
test the collective correlations arising in the \textit{ab initio} description
and to establish predictive power for unknown electromagnetic observables.

However, obtaining such predictions can be challenging, for reasons which vary
depending upon the many-body
method~\cite{pervin2007:qmc-matrix-elements-a6-7,maris2013:ncsm-pshell,carlson2015:qmc-nuclear,odell2016:ir-extrap-quadrupole}.
In particular, in \textit{ab initio} no-core configuration interaction (NCCI),
or no-core shell-model (NCSM), calculations~\cite{barrett2013:ncsm}, matrix
elements of the $E2$ operator are sensitive to the large-distance tails of the
nuclear wave function, which converge slowly in an oscillator-basis expansion.
It becomes computationally prohibitive to include the basis configurations
needed to obtain results of sufficient accuracy.  Other ``long-range''
observables, such as root mean square (r.m.s.)\ radii~\cite{bogner2008:ncsm-converg-2N}, exhibit
similarly challenging convergence properties.

Even in the face of such delayed convergence, useful predictions for $E2$
observables have been extracted, by focusing not on the absolute scale of
individual $E2$ matrix elements, but rather on their ratios.  It is found
empirically that the truncation error introduced by working in a
finite-dimensional many-body space is correlated between different $E2$ matrix
elements among low-lying states sharing similar structure (\textit{e.g.},
members of low-lying rotational
bands~\cite{caprio2013:berotor,*maris2015:berotor2,*maris2019:berotor2-ERRATUM,calci2016:observable-correlations-chiral,caprio2019:bebands-sdanca19,*caprio2022:10be-shape-sdanca21,caprio2020:bebands}
or isobaric analog states~\cite{henderson2019:7be-coulex,caprio2021:emratio}).
The truncation error systematically cancels in the ratio of matrix elements, as
a shared error in normalization, and rapidly-converging predictions are thus
obtained for such ratios.

This observation also provides an indirect route to predictions of absolute
strengths.  One may calibrate the absolute scale of calculated $E2$ observables
to a single well-measured $E2$
observable~\cite{calci2016:observable-correlations-chiral,caprio:8li-trans-PREPRINT},
\textit{e.g.}, the ground-state quadrupole moment, a property which is
well-measured for many nuclei~\cite{stone2016:e2-moments}.  (Such an approach
applies likewise to weak-interaction recoil-order form
factors~\cite{sargsyan2022:8li-clustering-beta-recoil}, which involve a
similar operator structure.)

In this letter, we demonstrate that, furthermore, the convergence of $E2$ matrix
elements is strongly correlated to that of electric monopole ($E0$) moments or,
equivalently, r.m.s.\ radius observables, in NCCI calculations.  Therefore,
robust, quantitative \textit{ab initio} predictions for the absolute scale of
$E2$ observables may be made by calibrating to the experimentally-known
ground-state charge radius, an observable which is known to exquisite precision
for a large subset of nuclei~\cite{angeli2013:charge-radii}, and which (unlike the quadrupole moment) is not
subject to selection rules on the ground-state angular momentum.  We first lay
out the expected relations between $E2$ and radius observables, in terms of
dimensionless ratios.  We then demonstrate the robust convergence obtained for
these ratios, and compare the resulting predictions against experiment.  For
purposes of illustration, we take $E2$ transition strengths in $\isotope[7]{Be}$
and $\isotope[10]{Be}$ and a selection of quadrupole moments in the lower $p$
shell.

 \paragraph*{Dimensionless ratios.}
Both the $E2$ transition strength and $E2$ moment are defined in terms of matrix elements of the
$E2$ operator, 
\begin{math}
Q_{2\mu}=\sum_{i\in p}e
r_{i}^2Y_{2\mu}(\uvec{r}_{i}),
\end{math}
where the summation runs over the (charged) protons.  Writing both observables in
terms of reduced matrix elements, to highlight the relationship,
\begin{math}
B(E2;J_i\rightarrow J_f)
\propto \abs{\trme{J_f}{
\,\sum_{i\in p} e r_{i}^2Y_{2}(\uvec{r}_{i})\,
}{J_i}}^2
\end{math}
and
\begin{math}
eQ(J)
\propto\trme{J}{
\,\sum_{i\in p} e r_{i}^2Y_{2}(\uvec{r}_{i})\,
}{J}.
\end{math}
Thus, among the $E2$ 
observables, $B(E2)/(eQ)^2$ is dimensionless, and involves a ratio of matrix elements of the form
\begin{displaymath}
{
\trme{\cdots}{
\,
\sum_{i\in p} r_{i}^2Y_{2}(\uvec{r}_{i})
\,
}{\cdots}
}
/
{
\trme{\cdots}{
\,
\sum_{i\in p} r_{i}^2Y_{2}(\uvec{r}_{i})
\,
}{\cdots}
},
\end{displaymath}
suggesting that truncation errors may cancel, as has been exploited in previous
work~\cite{calci2016:observable-correlations-chiral,caprio:8li-trans-PREPRINT}.

Then, the r.m.s.\ point-proton radius $r_p$ may be extracted from the
experimentally-observable charge radius
$r_c$~\cite{friar1997:charge-radius-correction,lu2013:laser-neutron-rich}.  This
$r_p$ is defined in terms of the monopole moment $M_0=\tme{JM}{\sum_{i\in p}
r_{i}^2}{JM}$ (independent of $M$) by
\begin{math}
r_p=(M_0/Z)^{1/2}.
\end{math}
Thus, again, we have an observable which is proportional to the matrix
element of a one-body operator with an $r^2$ radial dependence.  In terms of a reduced matrix element, 
\begin{math}
r_p^2
\propto \trme{J}{
\,\sum_{i\in p}r_{i}^2\,
}{J}.
\end{math}
Thus, ratios $B(E2)/(e^2r_p^4)$ or $Q/r_p^2$ are dimensionless, and
involve ratios of matrix elements of the form
\begin{displaymath}
{
\trme{\cdots}{
\,
\sum_{i\in p} r_{i}^2Y_{2}(\uvec{r}_{i})
\,
}{\cdots}
}
/
{
\trme{\cdots}{
\,
\sum_{i\in p} r_{i}^2
\,
}{\cdots}
}.
\end{displaymath}
To the extent that the truncation errors arising in NCCI calculations for such
matrix elements arise from omission of the tails of the wave functions, which
are subjected to the same $r^2$ weighting in either matrix element, it is not
unreasonable to anticipate that errors might again cancel in the ratio.
 \begin{figure}
\begin{center}
\includegraphics[width=\ifproofpre{1}{0.65}\hsize]{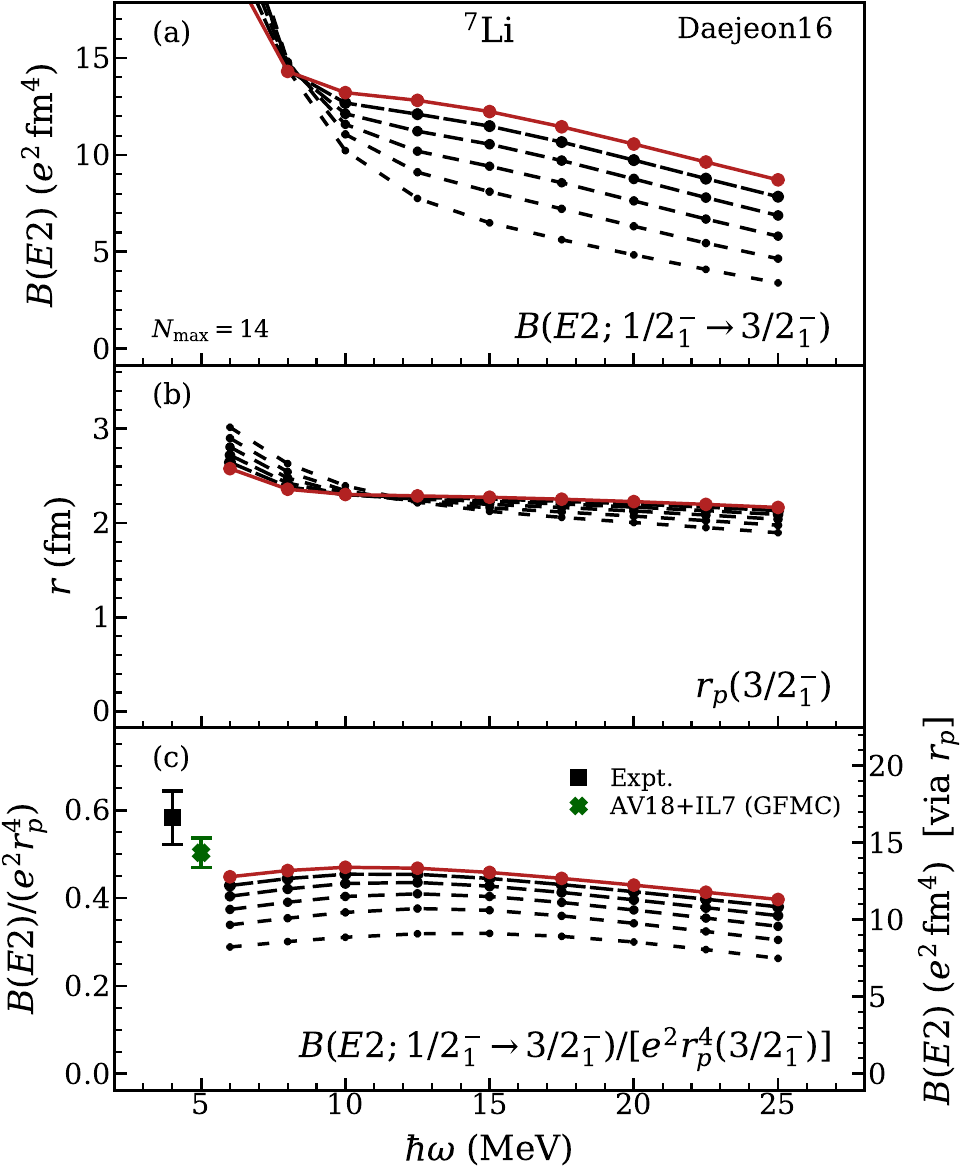}
\end{center}
\caption{Calculated (a)~$B(E2; 1/2^-_1\rightarrow 3/2^-_1)$,
  (b)~$r_p(3/2^-_1)$, and (c)~ratio
  $B(E2)/(e^2r_p^4)$, for $\isotope[7]{Li}$.  When calibrated to the experimental value
  for $r_p$, this ratio provides a prediction for the absolute $B(E2)$ (scale at
  right).  Calculated values obtained with the Daejeon16 interactions are shown as functions of
  the basis parameter $\hw$, from $\Nmax=4$ (short dashed curves) to $14$
  (solid curves).  For comparison, the experimental result~\cite{npa2002:005-007} (square) and GFMC AV18+IL7
  prediction~\cite{pastore2013:qmc-em-alt9} (cross) are shown.
\label{fig:be2-norm-rp-scan-7li}
}
\end{figure}

\begin{figure}
\begin{center}
\includegraphics[width=\ifproofpre{1}{0.75}\hsize]{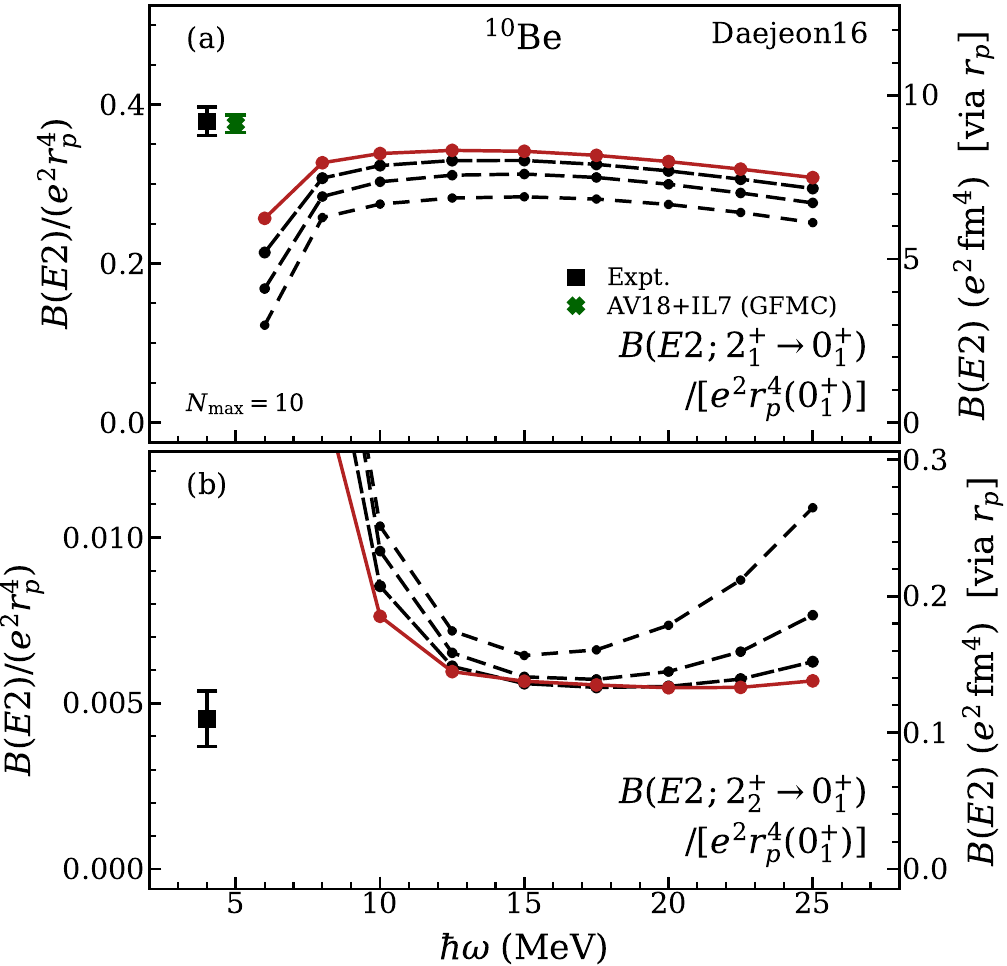}
\end{center}
\caption{ Calculated (a)~$B(E2; 2^+_1\rightarrow 0^+_1)$ and (b)~$B(E2;
  2^+_2\rightarrow 0^+_1)$ in $\isotope[10]{Be}$, expressed as ratios to
  $e^2r_p(0^+_1)^4$.  When calibrated to the experimental value for $r_p$, these
  ratios provide predictions for the absolute $B(E2)$ (see scale at right).
  Calculated values obtained with the Daejeon16 interaction are shown as
  functions of the basis parameter $\hw$, from $\Nmax=4$ (short dashed curves)
  to $10$ (solid curves).  For comparison, experimental
  results~\cite{mccutchan2009:10be-dsam-gfmc} (squares) and GFMC AV18+IL7
  predictions~\cite{mccutchan2009:10be-dsam-gfmc,mccutchan2012:10c-dsam,carlson2015:qmc-nuclear}
  (crosses) [off-scale in~(b)] are shown.
  \label{fig:be2-norm-rp-scan-10be}
}
\end{figure}

\paragraph*{Transitions.}
We consider first the $E2$ strength between the $1/2^-$ excited and
$3/2^-$ ground states of $\isotope[7]{Li}$, known experimentally to be
$B(E2;1/2^-_1\rightarrow 3/2^-_1)=16.6(10)\,e^2\fm^4$~\cite{npa2002:005-007}, from
Coulomb excitation~\cite{weller1985:7li-coulex}.  These states are commonly
interpreted as members of a $K=1/2$ cluster molecular rotational band (the
inverted ordering arising from Coriolis
decoupling~\cite{rowe2010:collective-motion}), hence the enhanced $E2$ strength.
The known ground-state charge radius of $\isotope[7]{Li}$~\cite{angeli2013:charge-radii}
gives $r_p=2.31(5)\,\fm$.

The convergence of the calculated $B(E2)$, with respect
to NCCI basis parameters, is seen in Fig.~\ref{fig:be2-norm-rp-scan-7li}(a).
These results are from calculations with the Daejeon16 internucleon
interaction~\cite{shirokov2016:nn-daejeon16}, carried out with the code
MFDn~\cite{aktulga2014:mfdn-scalability,shao2018:ncci-preconditioned}.  The
basis is truncated by restriction to configurations with some maximum
number $\Nmax$~\cite{barrett2013:ncsm} of oscillator excitations, relative to the lowest
Pauli-allowed configuration.  Moreover, the space spanned by these
configurations depends upon the oscillator length for the harmonic
oscillator orbitals, conventionally stated in terms of the
oscillator energy $\hw$~\cite{suhonen2007:nucleons-nucleus}.  Each curve in
Fig.~\ref{fig:be2-norm-rp-scan-7li} represents the results of calculations
sharing the same $\Nmax$ (from $4$ to $14$), for varying $\hw$.  

An approach to the true result, \textit{i.e.}, that which would be obtained from
solution of the many-body problem in an untruncated space, is signaled by a
value which no longer changes with increasing $\Nmax$ (compression of successive
curves) and is locally insensitive to the choice of basis length scale (flatness
or ``shouldering'' with respect to $\hw$).  While the curves in
Fig.~\ref{fig:be2-norm-rp-scan-7li}(a) may show some such tendencies, neither of
these signatures of convergence is sufficiently developed for us to read off a
concrete estimate of the result for the full, untruncated space.

The convergence of the calculated radius, $r_p(3/2^-_1)$, is likewise shown in
Fig.~\ref{fig:be2-norm-rp-scan-7li}(b).  While the $\hw$-dependence is
superficially less pronounced than for the $B(E2)$
[Fig.~\ref{fig:be2-norm-rp-scan-7li}(a)], recall that the radius goes as the
square root of the matrix element of an operator with $r^2$ radial dependence,
while the $B(E2)$ goes as the square of such a matrix element, and higher powers
amplify relative changes.  The $\hw$ dependence is qualitatively similar for
both observables, in that the values obtained at lower $\Nmax$ rise towards
infinity at small $\hw$ and fall towards zero at large $\hw$.  This behavior is
dictated by the scaling $b\propto (\hw)^{-1/2}$ of the oscillator length of the
underlying single-particle basis functions~\cite{suhonen2007:nucleons-nucleus}.

A more significant qualitative difference is that the curves representing $r_p$
for different $\Nmax$ cross each other~\cite{bogner2008:ncsm-converg-2N} for
$\hw$ between $\approx10\,\MeV$ and $15\,\MeV$, while those for the $B(E2)$ do
so only for lower $\hw$ and in a region of much more rapid change.  This makes
it clear that the two observables are not strictly proportional.

Nonetheless, we find that taking the appropriate dimensionless ratio,
$B(E2)/(e^2r_p^4)$, as shown in Fig.~\ref{fig:be2-norm-rp-scan-7li}(c), tames
the $\hw$-dependence of the calculated $B(E2;1/2^-_1\rightarrow 3/2^-_1)$.
Moreover, the spacing between curves for successive $\Nmax$ decreases
systematically, by very roughly a factor of $2$ with each step in $\Nmax$,
suggesting a geometric progression towards a converged value.  Such compression
is at best hinted in the underlying $B(E2)$ calculations
[Fig.~\ref{fig:be2-norm-rp-scan-7li}(a)], where it is rendered less relevant by
the confounding $\hw$ dependence.

Calibrating to the known radius gives the scale shown at right in
Fig.~\ref{fig:be2-norm-rp-scan-7li}(c).  An estimated ratio of
$B(E2;1/2^-_1\rightarrow 3/2^-_1)/[e^2r_p(3/2^-_1)^4]\approx0.50$ yields
$B(E2;1/2^-_1\rightarrow 3/2^-_1)\approx14\,e^2\fm^4$.  This is at the lower edge of
the uncertainties on the experimental value $0.58(6)$ for this ratio [Fig.~\ref{fig:be2-norm-rp-scan-7li}(c) (square)].

The \textit{ab initio} Green's function Monte Carlo
(GFMC)~\cite{carlson2015:qmc-nuclear} approach also yields predictions for $E2$
and radius observables for lower $p$-shell nuclei, which provide a theoretical
point of comparison.  The predicted $B(E2)/(e^2r_p^4)$ from GFMC
calculations~\cite{pastore2013:qmc-em-alt9}, with the Argonne $v_{18}$ (AV18)
two-nucleon~\cite{wiringa1995:nn-av18} and Illinois-7 (IL7)
three-nucleon~\cite{pieper2008:3n-il7-fm50} potentials, is shown
[Fig.~\ref{fig:be2-norm-rp-scan-7li}(c) (cross)].  Here the dominant
uncertainties are statistical in nature, and errors would not be expected to
cancel in the ratio, which is only taken for purposes of comparison (the
uncertainty in the $E2$ strength dominates that of the ratio).  The NCCI
Daejeon16 results are converging, with increasing $\Nmax$, in the direction of
the GFMC AV18+IL7 result, and the calculated value at the highest $\Nmax$ is
already consistent with the GFMC result to within statistical uncertainties.

Let us turn now to $\isotope[10]{Be}$, for which we consider the
$2^+_1\rightarrow 0^+_1$ transition, within the ground-state ($K=0$) rotational
band, and the $2^+_2\rightarrow 0^+_1$ transition, which is understood as an
interband transition from a proposed $K=2$ side band to the ground state
band~\cite{kanadaenyo1999:10be-amd,bohlen2007:10be-pickup,caprio2019:bebands-sdanca19,*caprio2022:10be-shape-sdanca21}.
(Intriguingly, the low-lying states may have a proton-neutron asymmetric
triaxial
deformation~\cite{kanadaenyo1997:c-amd-pn-decoupling,suhara2010:amd-deformation},
so that these bands together form a triaxial rotational
spectrum~\cite{davydov1958:arm-intro,meyertervehn1975:triax-odda}.)  The
$2^+\rightarrow 0^+$ transition strength within the ground-state band of
$\isotope[10]{Be}$ was known from early Doppler-shift lifetime
measurements~\cite{warburton1966:a10-a11-a12-dsam,fisher1968:a10-dsam}, but a
more recent experiment refines $B(E2;2^+_1\rightarrow0^+_1)$
from $10.5(10)\,e^2\fm^4$~\cite{raman2001:systematics} to
$9.2(3)\,e^2\fm^4$~\cite{mccutchan2009:10be-dsam-gfmc}, while the newly-measured
$2^+_2$ lifetime gives
$B(E2;2^+_2\rightarrow0^+_1)=0.11(2)\,e^2\fm^4$~\cite{mccutchan2009:10be-dsam-gfmc}.

The NCCI calculations for the dimensionless ratio $B(E2)/(e^2r_p^4)$, for each
of these transitions, is shown in Fig.~\ref{fig:be2-norm-rp-scan-10be}.  The
corresponding $B(E2)$, calibrated to the known ground-state
$r_p=2.22(2)\,\fm$~\cite{angeli2013:charge-radii}, is given by the scale at
right.

The ratio for the in-band transition [Fig.~\ref{fig:be2-norm-rp-scan-10be}(a)]
converges steadily from below, much as for the $\isotope[7]{Li}$ transition
[Fig.~\ref{fig:be2-norm-rp-scan-7li}(c)].  We may read off an estimated ratio of
$\approx 0.35$, which gives $B(E2;2^+_1\rightarrow0^+_1)\approx9\,e^2\fm^4$,
consistent with experiment~\cite{mccutchan2009:10be-dsam-gfmc}
[Fig.~\ref{fig:be2-norm-rp-scan-10be}(a) (square)].

The ratio for the interband transition [Fig.~\ref{fig:be2-norm-rp-scan-10be}(b)]
has a more dramatic $\hw$-dependence at low $\Nmax$, but the curves rapidly
compress and flatten for $\Nmax\gtrsim8$.  An estimated ratio of $\approx
0.005\text{--}0.006$ gives
$B(E2;2^+_2\rightarrow0^+_1)\approx0.12\text{--}0.14\,e^2\fm^4$.  Thus,
remarkably, the predicted interband transition strength is consistent with
experiment [Fig.~\ref{fig:be2-norm-rp-scan-10be}(b) (square)], to within uncertainties, even though this strength is nearly two
orders of magnitude weaker than the in-band strength (and an order of magnitude
weaker than the Weisskopf single-particle estimate).
While the GFMC AV18+IL7
prediction~\cite{mccutchan2009:10be-dsam-gfmc,mccutchan2012:10c-dsam,carlson2015:qmc-nuclear}
for the in-band transition [Fig.~\ref{fig:be2-norm-rp-scan-10be}(a) (cross)] is
in close agreement with the present results and consistent with experiment, the
prediction for the weaker interband transition, at $1.7(1)\,e^2\fm^4$, lies off
scale in Fig.~\ref{fig:be2-norm-rp-scan-10be}(b).

 \begin{figure*}
\begin{center}
\includegraphics[width=\ifproofpre{0.75}{1}\hsize]{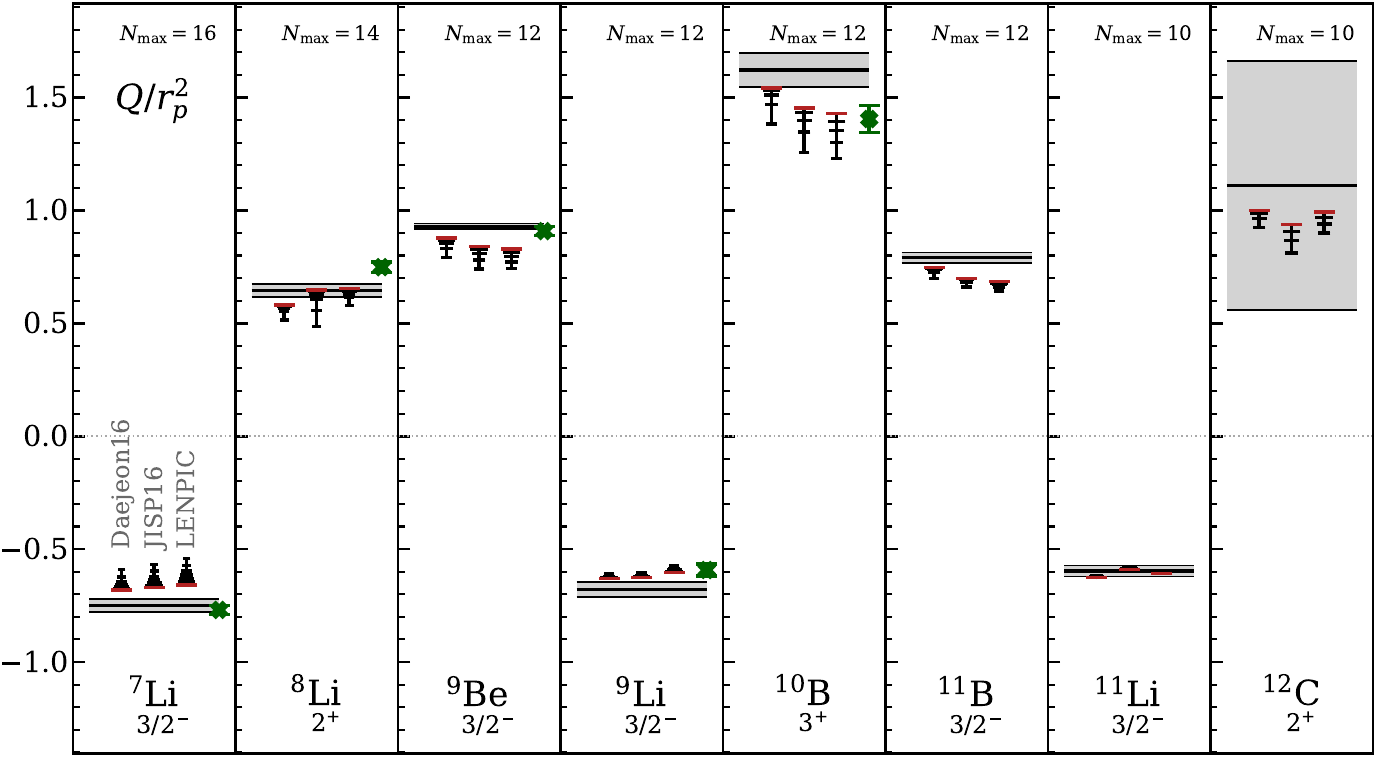}
\end{center}
\caption{Calculated ground-state quadrupole moment, normalized to the proton
  radius, for nuclei ($7\leq A \leq 11$) where both quantities are
  experimentally known.  The $\isotope[12]{C}$ $2^+$ excited-state quadrupole
  moment, normalized to the ground-state proton radius, is also shown.
  Predictions are obtained with the Daejeon16, JISP16, and LENPIC interactions
  (from left to right, for each nucleus).  Calculated values are shown at fixed
  $\hw$ ($15\,\MeV$, $20\,\MeV$, and $25\,\MeV$, respectively, for the three
  interactions), from $\Nmax=4$ to the maximum $\Nmax$ indicated (at top).  For
  comparison, the experimental results~\cite{stone2016:e2-moments} are shown
  (horizontal line and error band, where the signs of some quadrupole moments
  are experimentally undetermined), as are the GFMC AV18+IL7
  predictions~\cite{pastore2013:qmc-em-alt9,pieper:cited} (crosses).
\label{fig:e2-moment-norm-rp-teardrop}
}
\end{figure*}

\paragraph*{Moments.}
We examine the dimensionless ratio $Q/r_p^2$, for a selection of nuclei from the
lower $p$ shell, in Fig.~\ref{fig:e2-moment-norm-rp-teardrop}.  Ground-state
electric quadrupole moments are experimentally known~\cite{stone2016:e2-moments}
to comparatively high precision ($\approx1\text{--}10\%$) for many of the
$p$-shell nuclei having ground-state angular momenta $J\geq1$, and ground-state
charge radii are known~\cite{angeli2013:charge-radii} for most of the stable and
neutron-rich $p$-shell nuclei.  We thus have grounds for stringent tests of the
ability of NCCI calculations to predict the dimensionless ratio $Q/r_p^2$.

For conciseness, in Fig.~\ref{fig:e2-moment-norm-rp-teardrop}, we show only the
$\Nmax$ dependence of calculated results at a fixed $\hw$, which is chosen as the
approximate location of the variational energy minimum.  (For detailed
illustrations of the convergence for some of these quadrupole moments, see
Ref.~\cite{caprio2021:emratio}.)  All nuclei for which both the ground-state
quadrupole moment and charge radius are known, for $7\leq A \leq 11$, are
included.

While the results discussed thus far
(Figs.~\ref{fig:be2-norm-rp-scan-7li}--\ref{fig:be2-norm-rp-scan-10be}) have
been based on calculations using the Daejeon16 interaction, interactions with
notably different convergence rates for the underlying moment or radius still
yield well-behaved convergence for the dimensionless ratio.  In
Fig.~\ref{fig:e2-moment-norm-rp-teardrop}, we provide comparison with results
obtained with the JISP16 $J$-matrix inverse scattering
interaction~\cite{shirokov2007:nn-jisp16} and the unsoftened LENPIC chiral EFT
interaction (specifically, the two-body part at \ntwolo{}, using a semi-local
coordinate-space regulator with
$R=1\,\fm$)~\cite{epelbaum2015:lenpic-n4lo-scs,epelbaum2015:lenpic-n3lo-scs},
shown from left to right within each panel of
Fig.~\ref{fig:e2-moment-norm-rp-teardrop}.

We see a uniformly rapid approach to convergence for the ratio $Q/r_p^2$, across
nuclei and interactions, as evidenced by the sizes of successive steps in the
computed values, which decrease in a roughly geometric fashion with successive
steps in $\Nmax$.  We thus have the means to meaningfully estimate the true
value for this ratio, for the given interaction, in the
untruncated many-body space.  The predictions for $Q/r_p^2$ are not strongly
dependent on the interaction, with differences at the $\lesssim 10\%$ level (it
must be kept in mind that some of the apparent differences in
Fig.~\ref{fig:e2-moment-norm-rp-teardrop} may simply reflect the
still-incomplete convergence of the results).

Comparing to the experimental ratios in
Fig.~\ref{fig:e2-moment-norm-rp-teardrop} (horizontal lines and error bands), we
see that the NCCI predictions are consistent with experiment to within
$\approx10\%$ in all cases.  The GFMC AV18+IL7 predictions for
$A\leq9$~\cite{pastore2013:qmc-em-alt9} are also shown [Fig.~\ref{fig:e2-moment-norm-rp-teardrop} (crosses)].  A notable
discrepancy between the present predictions and experiment arises
for $\isotope[7]{Li}$, where a robust prediction is obtained for the ratio
(reasonably well-converged with $\Nmax$ and independent of interaction), roughly
$10\%$ smaller in magnitude than the experimental value, and outside the
uncertainties by a factor of $\approx 2$ (in contrast, the GFMC AV18+IL7
predictions are consistent with experiment).  For $\isotope[11]{Li}$, the rapid
convergence of the results, along with their interaction-independence and
agreement with experiment to within relatively narrow ($\approx 2\%$)
uncertainties, is particularly notable, considering the challenging neutron halo
structure of this nuclide~\cite{tanihata1985:radii-11li-halo,jonson2004:light-dripline}.

Finally, the quadrupole moment for the first excited $2^+$ state in
$\isotope[12]{C}$ is experimentally known~\cite{stone2016:e2-moments}, though to
lower precision than the ground-state moments, from the reorientation effect in
Coulomb excitation~\cite{vermeer1983:12c-coulex}, as is the charge
radius~\cite{angeli2013:charge-radii} of the $0^+$ ground state.  Here, we still
take the ratio of an electric quadruple moment and a proton radius, but now of
distinct (though structurally similar) states within the $\isotope[12]{C}$
ground state rotational band.  The calculated $Q(2^+_1)/r_p(0^+_1)^2$ is shown in
Fig.~\ref{fig:e2-moment-norm-rp-teardrop}~(at far right).  Taking this ratio
again provides rapid convergence with $\Nmax$, yielding a result which not only
is consistent with experiment but also offers a prediction of higher precision.
 \paragraph*{Conclusion.}
\textit{Ab initio} predictions of nuclear
$E2$ observables are hampered by sensitivity to the large-distance behavior of
the nuclear wave function, resulting in poor convergence in NCCI calculations.
However, we demonstrate that, when calculated $E2$ observables are normalized to
the calculated radius, taken in the appropriate power to generate a
dimensionless ratio, systematic truncation errors cancel, and comparatively
rapid convergence is obtained.

Since nuclear ground state charge radii are well-measured for an appreciable
subset of nuclei, robust \textit{ab initio} predictions of ratios
$B(E2)/(e^2r_p^4)$ or $Q/r_p^2$ effectively yield predictions of the $E2$
strengths or quadrupole moments themselves.  The present approach is doubtless
subject to limitations in its applicability, whether for $E2$ observables
involving excited states with significantly different structure from the ground
state providing the normalization, or in cases where where convergence is
governed by delicate sensitivity to mixing.  Nonetheless, as we have illustrated
for a range of transition strengths and moments in $p$-shell nuclei, meaningful
\textit{ab initio} NCCI predictions for $E2$
observables can be obtained by normalizing to the experimentally-known charge
radius.

Ideally, one may seek to directly improve convergence of $E2$ (and radius)
observables in NCCI calculations, \textit{e.g.}, through explicit inclusion of
clustering degrees of freedom (as in the no-core shell model with
continuum~\cite{navratil2016:ncsmc,vorabbi2019:7be-7li-ncsmc}), implicit
inclusion of giant quadrupole resonance degrees of freedom via
$\grpsptr$ symmetry-adapted
calculations~\cite{rowe1985:micro-collective-sp6r,dytrych2008:sp-ncsm,mccoy2020:spfamilies},
improved asymptotics through transformation to an alternative single-particle
basis~\cite{constantinou2017:natorb-natowitz16,*fasano2022:natorb}, or
perturbative importance truncation schemes~\cite{roth2007:it-ncsm-40ca}.
Dimensionless ratios of the type presented may be used in combination with such
approaches to boosting convergence, as has already been illustrated for ratios of $E2$
observables taken in conjunction with importance
truncation~\cite{calci2016:observable-correlations-chiral}.

\begin{acknowledgments}
We thank James P.~Vary, Ik Jae Shin, and Youngman Kim for sharing illuminating
results on ratios of observables, Augusto O.~Macchiavelli for valuable
discussions, and Jakub Herko and Zhou Zhou for comments on the manuscript.  This
material is based upon work supported by the U.S.~Department of Energy, Office
of Science, under Awards No.~DE-FG02-95ER40934 and DE-SC0018223
(SciDAC4/NUCLEI).  An award of computer time was provided by the Innovative and
Novel Computational Impact on Theory and Experiment (INCITE) program. This
research used computational resources of the National Energy Research Scientific
Computing Center (NERSC) and the Argonne Leadership Computing Facility (ALCF),
which are U.S.~Department of Energy, Office of Science, user facilities,
supported under Contracts No.~DE-AC02-05CH11231 and DE-AC02-06CH11357.
\end{acknowledgments}

\bibliographystyle{apsrev4-2}

\begin{thebibliography}{68}\makeatletter
\providecommand \@ifxundefined [1]{\@ifx{#1\undefined}
}\providecommand \@ifnum [1]{\ifnum #1\expandafter \@firstoftwo
 \else \expandafter \@secondoftwo
 \fi
}\providecommand \@ifx [1]{\ifx #1\expandafter \@firstoftwo
 \else \expandafter \@secondoftwo
 \fi
}\providecommand \natexlab [1]{#1}\providecommand \enquote  [1]{``#1''}\providecommand \bibnamefont  [1]{#1}\providecommand \bibfnamefont [1]{#1}\providecommand \citenamefont [1]{#1}\providecommand \href@noop [0]{\@secondoftwo}\providecommand \href [0]{\begingroup \@sanitize@url \@href}\providecommand \@href[1]{\@@startlink{#1}\@@href}\providecommand \@@href[1]{\endgroup#1\@@endlink}\providecommand \@sanitize@url [0]{\catcode `\\12\catcode `\$12\catcode
  `\&12\catcode `\#12\catcode `\^12\catcode `\_12\catcode `\%12\relax}\providecommand \@@startlink[1]{}\providecommand \@@endlink[0]{}\providecommand \url  [0]{\begingroup\@sanitize@url \@url }\providecommand \@url [1]{\endgroup\@href {#1}{\urlprefix }}\providecommand \urlprefix  [0]{URL }\providecommand \Eprint [0]{\href }\providecommand \doibase [0]{https://doi.org/}\providecommand \selectlanguage [0]{\@gobble}\providecommand \bibinfo  [0]{\@secondoftwo}\providecommand \bibfield  [0]{\@secondoftwo}\providecommand \translation [1]{[#1]}\providecommand \BibitemOpen [0]{}\providecommand \bibitemStop [0]{}\providecommand \bibitemNoStop [0]{.\EOS\space}\providecommand \EOS [0]{\spacefactor3000\relax}\providecommand \BibitemShut  [1]{\csname bibitem#1\endcsname}\let\auto@bib@innerbib\@empty
\bibitem [{\citenamefont {Bohr}\ and\ \citenamefont
  {Mottelson}(1998)}]{bohr1998:v2}\BibitemOpen
  \bibfield  {author} {\bibinfo {author} {\bibfnamefont {A.}~\bibnamefont
  {Bohr}}\ and\ \bibinfo {author} {\bibfnamefont {B.~R.}\ \bibnamefont
  {Mottelson}},\ }\href@noop {} {\emph {\bibinfo {title} {Nuclear
  Structure}}},\ Vol.~\bibinfo {volume} {2}\ (\bibinfo  {publisher} {World
  Scientific},\ \bibinfo {address} {Singapore},\ \bibinfo {year}
  {1998})\BibitemShut {NoStop}\bibitem [{\citenamefont {Casten}(2000)}]{casten2000:ns}\BibitemOpen
  \bibfield  {author} {\bibinfo {author} {\bibfnamefont {R.~F.}\ \bibnamefont
  {Casten}},\ }\href@noop {} {\emph {\bibinfo {title} {Nuclear Structure from a
  Simple Perspective}}},\ \bibinfo {edition} {2nd}\ ed.,\ \bibinfo {series}
  {Oxford Studies in Nuclear Physics}\ No.~\bibinfo {number} {23}\ (\bibinfo
  {publisher} {Oxford University Press},\ \bibinfo {address} {Oxford},\
  \bibinfo {year} {2000})\BibitemShut {NoStop}\bibitem [{\citenamefont {Rowe}(2010)}]{rowe2010:collective-motion}\BibitemOpen
  \bibfield  {author} {\bibinfo {author} {\bibfnamefont {D.~J.}\ \bibnamefont
  {Rowe}},\ }\href@noop {} {\emph {\bibinfo {title} {Nuclear Collective Motion:
  Models and Theory}}}\ (\bibinfo  {publisher} {World Scientific},\ \bibinfo
  {address} {Singapore},\ \bibinfo {year} {2010})\BibitemShut {NoStop}\bibitem [{\citenamefont {Raman}\ \emph {et~al.}(2001)\citenamefont {Raman},
  \citenamefont {Nestor},\ and\ \citenamefont
  {Tikkanen}}]{raman2001:systematics}\BibitemOpen
  \bibfield  {author} {\bibinfo {author} {\bibfnamefont {S.}~\bibnamefont
  {Raman}}, \bibinfo {author} {\bibfnamefont {C.~W.}\ \bibnamefont {Nestor},
  \bibfnamefont {Jr.}},\ and\ \bibinfo {author} {\bibfnamefont
  {P.}~\bibnamefont {Tikkanen}},\ }\bibfield  {title} {\bibinfo {title}
  {Transition probability from the ground to the first-excited $2^+$ state of
  even–even nuclides},\ }\href {https://doi.org/10.1006/adnd.2001.0858}
  {\bibfield  {journal} {\bibinfo  {journal} {At. Data Nucl. Data Tables}\
  }\textbf {\bibinfo {volume} {78}},\ \bibinfo {pages} {1} (\bibinfo {year}
  {2001})}\BibitemShut {NoStop}\bibitem [{\citenamefont {Alaga}\ \emph {et~al.}(1955)\citenamefont {Alaga},
  \citenamefont {Alder}, \citenamefont {Bohr},\ and\ \citenamefont
  {Mottelson}}]{alaga1955:branching}\BibitemOpen
  \bibfield  {author} {\bibinfo {author} {\bibfnamefont {G.}~\bibnamefont
  {Alaga}}, \bibinfo {author} {\bibfnamefont {K.}~\bibnamefont {Alder}},
  \bibinfo {author} {\bibfnamefont {A.}~\bibnamefont {Bohr}},\ and\ \bibinfo
  {author} {\bibfnamefont {B.~R.}\ \bibnamefont {Mottelson}},\ }\bibfield
  {title} {\bibinfo {title} {Intensity rules for beta and gamma transitions to
  nuclear rotational states},\ }\href
  {http://gymarkiv.sdu.dk/MFM/kdvs/mfm\%2020-29/mfm-29-9.pdf} {\bibfield
  {journal} {\bibinfo  {journal} {Mat. Fys. Medd. Dan. Vid. Selsk.}\ }\textbf
  {\bibinfo {volume} {29}}(9) (\bibinfo {year} {1955})}\BibitemShut {NoStop}\bibitem [{\citenamefont {Pieper}\ \emph {et~al.}(2004)\citenamefont {Pieper},
  \citenamefont {Wiringa},\ and\ \citenamefont
  {Carlson}}]{pieper2004:gfmc-a6-8}\BibitemOpen
  \bibfield  {author} {\bibinfo {author} {\bibfnamefont {S.~C.}\ \bibnamefont
  {Pieper}}, \bibinfo {author} {\bibfnamefont {R.~B.}\ \bibnamefont
  {Wiringa}},\ and\ \bibinfo {author} {\bibfnamefont {J.}~\bibnamefont
  {Carlson}},\ }\bibfield  {title} {\bibinfo {title} {Quantum {M}onte {C}arlo
  calculations of excited states in {$A=6$--$8$} nuclei},\ }\href
  {https://doi.org/10.1103/PhysRevC.70.054325} {\bibfield  {journal} {\bibinfo
  {journal} {Phys. Rev. C}\ }\textbf {\bibinfo {volume} {70}},\ \bibinfo
  {pages} {054325} (\bibinfo {year} {2004})}\BibitemShut {NoStop}\bibitem [{\citenamefont {Neff}\ and\ \citenamefont
  {Feldmeier}(2004)}]{neff2004:cluster-fmd}\BibitemOpen
  \bibfield  {author} {\bibinfo {author} {\bibfnamefont {T.}~\bibnamefont
  {Neff}}\ and\ \bibinfo {author} {\bibfnamefont {H.}~\bibnamefont
  {Feldmeier}},\ }\bibfield  {title} {\bibinfo {title} {Cluster structures
  within fermionic molecular dynamics},\ }\href
  {https://doi.org/10.1016/j.nuclphysa.2004.04.061} {\bibfield  {journal}
  {\bibinfo  {journal} {Nucl. Phys. A}\ }\textbf {\bibinfo {volume} {738}},\
  \bibinfo {pages} {357} (\bibinfo {year} {2004})}\BibitemShut {NoStop}\bibitem [{\citenamefont {Maris}(2012)}]{maris2012:mfdn-ccp11}\BibitemOpen
  \bibfield  {author} {\bibinfo {author} {\bibfnamefont {P.}~\bibnamefont
  {Maris}},\ }\bibfield  {title} {\bibinfo {title} {\textit{Ab initio} nuclear
  structure calculations of light nuclei},\ }\href
  {https://doi.org/10.1088/1742-6596/402/1/012031} {\bibfield  {journal}
  {\bibinfo  {journal} {J. Phys. Conf. Ser.}\ }\textbf {\bibinfo {volume}
  {402}},\ \bibinfo {pages} {012031} (\bibinfo {year} {2012})}\BibitemShut
  {NoStop}\bibitem [{\citenamefont {Yoshida}\ \emph {et~al.}(2013)\citenamefont
  {Yoshida}, \citenamefont {Shimizu}, \citenamefont {Abe},\ and\ \citenamefont
  {Otsuka}}]{yoshida2013:ncmcsm-8be-10be-6be-cluster}\BibitemOpen
  \bibfield  {author} {\bibinfo {author} {\bibfnamefont {T.}~\bibnamefont
  {Yoshida}}, \bibinfo {author} {\bibfnamefont {N.}~\bibnamefont {Shimizu}},
  \bibinfo {author} {\bibfnamefont {T.}~\bibnamefont {Abe}},\ and\ \bibinfo
  {author} {\bibfnamefont {T.}~\bibnamefont {Otsuka}},\ }\bibfield  {title}
  {\bibinfo {title} {Intrinsic structure of light nuclei in {M}onte {C}arlo
  shell model calculation},\ }\href {https://doi.org/10.1007/s00601-013-0680-7}
  {\bibfield  {journal} {\bibinfo  {journal} {Few-Body Syst.}\ }\textbf
  {\bibinfo {volume} {54}},\ \bibinfo {pages} {1465} (\bibinfo {year}
  {2013})}\BibitemShut {NoStop}\bibitem [{\citenamefont {Romero-Redondo}\ \emph {et~al.}(2016)\citenamefont
  {Romero-Redondo}, \citenamefont {Quaglioni}, \citenamefont {Navr\'{a}til},\
  and\ \citenamefont {Hupin}}]{romeroredondo2016:6he-correlations}\BibitemOpen
  \bibfield  {author} {\bibinfo {author} {\bibfnamefont {C.}~\bibnamefont
  {Romero-Redondo}}, \bibinfo {author} {\bibfnamefont {S.}~\bibnamefont
  {Quaglioni}}, \bibinfo {author} {\bibfnamefont {P.}~\bibnamefont
  {Navr\'{a}til}},\ and\ \bibinfo {author} {\bibfnamefont {G.}~\bibnamefont
  {Hupin}},\ }\bibfield  {title} {\bibinfo {title} {How many-body correlations
  and $\alpha$ clustering shape $\isotope[6]{He}$},\ }\href
  {https://doi.org/10.1103/PhysRevLett.117.222501} {\bibfield  {journal}
  {\bibinfo  {journal} {Phys. Rev. Lett.}\ }\textbf {\bibinfo {volume} {117}},\
  \bibinfo {pages} {222501} (\bibinfo {year} {2016})}\BibitemShut {NoStop}\bibitem [{\citenamefont {Navr\'{a}til}\ \emph {et~al.}(2016)\citenamefont
  {Navr\'{a}til}, \citenamefont {Quaglioni}, \citenamefont {Hupin},
  \citenamefont {Romero-Redondo},\ and\ \citenamefont
  {Calci}}]{navratil2016:ncsmc}\BibitemOpen
  \bibfield  {author} {\bibinfo {author} {\bibfnamefont {P.}~\bibnamefont
  {Navr\'{a}til}}, \bibinfo {author} {\bibfnamefont {S.}~\bibnamefont
  {Quaglioni}}, \bibinfo {author} {\bibfnamefont {G.}~\bibnamefont {Hupin}},
  \bibinfo {author} {\bibfnamefont {C.}~\bibnamefont {Romero-Redondo}},\ and\
  \bibinfo {author} {\bibfnamefont {A.}~\bibnamefont {Calci}},\ }\bibfield
  {title} {\bibinfo {title} {Unified \textit{ab initio} approaches to nuclear
  structure},\ }\href {https://doi.org/10.1088/0031-8949/91/5/053002}
  {\bibfield  {journal} {\bibinfo  {journal} {Physica Scripta}\ }\textbf
  {\bibinfo {volume} {91}},\ \bibinfo {pages} {053002} (\bibinfo {year}
  {2016})}\BibitemShut {NoStop}\bibitem [{\citenamefont {Caprio}\ \emph {et~al.}(2013)\citenamefont {Caprio},
  \citenamefont {Maris},\ and\ \citenamefont {Vary}}]{caprio2013:berotor}\BibitemOpen
  \bibfield  {author} {\bibinfo {author} {\bibfnamefont {M.~A.}\ \bibnamefont
  {Caprio}}, \bibinfo {author} {\bibfnamefont {P.}~\bibnamefont {Maris}},\ and\
  \bibinfo {author} {\bibfnamefont {J.~P.}\ \bibnamefont {Vary}},\ }\bibfield
  {title} {\bibinfo {title} {Emergence of rotational bands in \textit{ab
  initio} no-core configuration interaction calculations of light nuclei},\
  }\href {https://doi.org/10.1016/j.physletb.2012.12.064} {\bibfield  {journal}
  {\bibinfo  {journal} {Phys. Lett. B}\ }\textbf {\bibinfo {volume} {719}},\
  \bibinfo {pages} {179} (\bibinfo {year} {2013})}\BibitemShut {NoStop}\bibitem [{\citenamefont {Maris}\ \emph {et~al.}(2015)\citenamefont {Maris},
  \citenamefont {Caprio},\ and\ \citenamefont {Vary}}]{maris2015:berotor2}\BibitemOpen
  \bibfield  {author} {\bibinfo {author} {\bibfnamefont {P.}~\bibnamefont
  {Maris}}, \bibinfo {author} {\bibfnamefont {M.~A.}\ \bibnamefont {Caprio}},\
  and\ \bibinfo {author} {\bibfnamefont {J.~P.}\ \bibnamefont {Vary}},\
  }\bibfield  {title} {\bibinfo {title} {Emergence of rotational bands in
  \textit{ab initio} no-core configuration interaction calculations of the
  $\isotope{Be}$ isotopes},\ }\href
  {https://doi.org/10.1103/PhysRevC.91.014310} {\bibfield  {journal} {\bibinfo
  {journal} {Phys. Rev. C}\ }\textbf {\bibinfo {volume} {91}},\ \bibinfo
  {pages} {014310} (\bibinfo {year} {2015})}\BibitemShut {NoStop}\bibitem [{\citenamefont {Maris}\ \emph {et~al.}(2019)\citenamefont {Maris},
  \citenamefont {Caprio},\ and\ \citenamefont
  {Vary}}]{maris2019:berotor2-ERRATUM}\BibitemOpen
  \bibfield  {author} {\bibinfo {author} {\bibfnamefont {P.}~\bibnamefont
  {Maris}}, \bibinfo {author} {\bibfnamefont {M.~A.}\ \bibnamefont {Caprio}},\
  and\ \bibinfo {author} {\bibfnamefont {J.~P.}\ \bibnamefont {Vary}},\
  }\bibfield  {title} {\bibinfo {title} {Erratum: Emergence of rotational bands
  in \textit{ab initio} no-core configuration interaction calculations of the
  $\isotope{Be}$ isotopes},\ }\href
  {https://doi.org/10.1103/PhysRevC.99.029902} {\bibfield  {journal} {\bibinfo
  {journal} {Phys. Rev. C}\ }\textbf {\bibinfo {volume} {99}},\ \bibinfo
  {pages} {029902(E)} (\bibinfo {year} {2019})}\BibitemShut {NoStop}\bibitem [{\citenamefont {Caprio}\ \emph {et~al.}(2015)\citenamefont {Caprio},
  \citenamefont {Maris}, \citenamefont {Vary},\ and\ \citenamefont
  {Smith}}]{caprio2015:berotor-ijmpe}\BibitemOpen
  \bibfield  {author} {\bibinfo {author} {\bibfnamefont {M.~A.}\ \bibnamefont
  {Caprio}}, \bibinfo {author} {\bibfnamefont {P.}~\bibnamefont {Maris}},
  \bibinfo {author} {\bibfnamefont {J.~P.}\ \bibnamefont {Vary}},\ and\
  \bibinfo {author} {\bibfnamefont {R.}~\bibnamefont {Smith}},\ }\bibfield
  {title} {\bibinfo {title} {Collective rotation from \textit{ab initio}
  theory},\ }\href {https://doi.org/10.1142/S0218301315410025} {\bibfield
  {journal} {\bibinfo  {journal} {Int. J. Mod. Phys. E}\ }\textbf {\bibinfo
  {volume} {24}},\ \bibinfo {pages} {1541002} (\bibinfo {year}
  {2015})}\BibitemShut {NoStop}\bibitem [{\citenamefont {Stroberg}\ \emph {et~al.}(2016)\citenamefont
  {Stroberg}, \citenamefont {Hergert}, \citenamefont {Holt}, \citenamefont
  {Bogner},\ and\ \citenamefont
  {Schwenk}}]{stroberg2016:ab-initio-sd-multireference}\BibitemOpen
  \bibfield  {author} {\bibinfo {author} {\bibfnamefont {S.~R.}\ \bibnamefont
  {Stroberg}}, \bibinfo {author} {\bibfnamefont {H.}~\bibnamefont {Hergert}},
  \bibinfo {author} {\bibfnamefont {J.~D.}\ \bibnamefont {Holt}}, \bibinfo
  {author} {\bibfnamefont {S.~K.}\ \bibnamefont {Bogner}},\ and\ \bibinfo
  {author} {\bibfnamefont {A.}~\bibnamefont {Schwenk}},\ }\bibfield  {title}
  {\bibinfo {title} {Ground and excited states of doubly open-shell nuclei from
  \textit{ab initio} valence-space {H}amiltonians},\ }\href
  {https://doi.org/10.1103/PhysRevC.93.051301} {\bibfield  {journal} {\bibinfo
  {journal} {Phys. Rev. C}\ }\textbf {\bibinfo {volume} {93}},\ \bibinfo
  {pages} {051301} (\bibinfo {year} {2016})}\BibitemShut {NoStop}\bibitem [{\citenamefont {Jansen}\ \emph {et~al.}(2016)\citenamefont {Jansen},
  \citenamefont {Schuster}, \citenamefont {Signoracci}, \citenamefont {Hagen},\
  and\ \citenamefont {Navr\'atil}}]{jansen2016:sd-shell-ab-initio}\BibitemOpen
  \bibfield  {author} {\bibinfo {author} {\bibfnamefont {G.~R.}\ \bibnamefont
  {Jansen}}, \bibinfo {author} {\bibfnamefont {M.~D.}\ \bibnamefont
  {Schuster}}, \bibinfo {author} {\bibfnamefont {A.}~\bibnamefont
  {Signoracci}}, \bibinfo {author} {\bibfnamefont {G.}~\bibnamefont {Hagen}},\
  and\ \bibinfo {author} {\bibfnamefont {P.}~\bibnamefont {Navr\'atil}},\
  }\bibfield  {title} {\bibinfo {title} {Open $sd$-shell nuclei from first
  principles},\ }\href {https://doi.org/10.1103/PhysRevC.94.011301} {\bibfield
  {journal} {\bibinfo  {journal} {Phys. Rev. C}\ }\textbf {\bibinfo {volume}
  {94}},\ \bibinfo {pages} {011301} (\bibinfo {year} {2016})}\BibitemShut
  {NoStop}\bibitem [{\citenamefont {Caprio}\ \emph {et~al.}(2020)\citenamefont {Caprio},
  \citenamefont {Fasano}, \citenamefont {Maris}, \citenamefont {McCoy},\ and\
  \citenamefont {Vary}}]{caprio2020:bebands}\BibitemOpen
  \bibfield  {author} {\bibinfo {author} {\bibfnamefont {M.~A.}\ \bibnamefont
  {Caprio}}, \bibinfo {author} {\bibfnamefont {P.~J.}\ \bibnamefont {Fasano}},
  \bibinfo {author} {\bibfnamefont {P.}~\bibnamefont {Maris}}, \bibinfo
  {author} {\bibfnamefont {A.~E.}\ \bibnamefont {McCoy}},\ and\ \bibinfo
  {author} {\bibfnamefont {J.~P.}\ \bibnamefont {Vary}},\ }\bibfield  {title}
  {\bibinfo {title} {Probing \textit{ab initio} emergence of nuclear
  rotation},\ }\href {https://doi.org/10.1140/epja/s10050-020-00112-0}
  {\bibfield  {journal} {\bibinfo  {journal} {Eur. Phys. J. A}\ }\textbf
  {\bibinfo {volume} {56}},\ \bibinfo {pages} {120} (\bibinfo {year}
  {2020})}\BibitemShut {NoStop}\bibitem [{\citenamefont {Pervin}\ \emph {et~al.}(2007)\citenamefont {Pervin},
  \citenamefont {Pieper},\ and\ \citenamefont
  {Wiringa}}]{pervin2007:qmc-matrix-elements-a6-7}\BibitemOpen
  \bibfield  {author} {\bibinfo {author} {\bibfnamefont {M.}~\bibnamefont
  {Pervin}}, \bibinfo {author} {\bibfnamefont {S.~C.}\ \bibnamefont {Pieper}},\
  and\ \bibinfo {author} {\bibfnamefont {R.~B.}\ \bibnamefont {Wiringa}},\
  }\bibfield  {title} {\bibinfo {title} {Quantum {M}onte {C}arlo calculations
  of electroweak transition matrix elements in {$A = 6$, $7$} nuclei},\ }\href
  {https://doi.org/10.1103/PhysRevC.76.064319} {\bibfield  {journal} {\bibinfo
  {journal} {Phys. Rev. C}\ }\textbf {\bibinfo {volume} {76}},\ \bibinfo
  {pages} {064319} (\bibinfo {year} {2007})}\BibitemShut {NoStop}\bibitem [{\citenamefont {Maris}\ and\ \citenamefont
  {Vary}(2013)}]{maris2013:ncsm-pshell}\BibitemOpen
  \bibfield  {author} {\bibinfo {author} {\bibfnamefont {P.}~\bibnamefont
  {Maris}}\ and\ \bibinfo {author} {\bibfnamefont {J.~P.}\ \bibnamefont
  {Vary}},\ }\bibfield  {title} {\bibinfo {title} {\textit{Ab initio} nuclear
  structure calculations of $p$-shell nuclei with {JISP16}},\ }\href
  {https://doi.org/10.1142/S0218301313300166} {\bibfield  {journal} {\bibinfo
  {journal} {Int. J. Mod. Phys. E}\ }\textbf {\bibinfo {volume} {22}},\
  \bibinfo {pages} {1330016} (\bibinfo {year} {2013})}\BibitemShut {NoStop}\bibitem [{\citenamefont {Carlson}\ \emph {et~al.}(2015)\citenamefont
  {Carlson}, \citenamefont {Gandolfi}, \citenamefont {Pederiva}, \citenamefont
  {Pieper}, \citenamefont {Schiavilla}, \citenamefont {Schmidt},\ and\
  \citenamefont {Wiringa}}]{carlson2015:qmc-nuclear}\BibitemOpen
  \bibfield  {author} {\bibinfo {author} {\bibfnamefont {J.}~\bibnamefont
  {Carlson}}, \bibinfo {author} {\bibfnamefont {S.}~\bibnamefont {Gandolfi}},
  \bibinfo {author} {\bibfnamefont {F.}~\bibnamefont {Pederiva}}, \bibinfo
  {author} {\bibfnamefont {S.~C.}\ \bibnamefont {Pieper}}, \bibinfo {author}
  {\bibfnamefont {R.}~\bibnamefont {Schiavilla}}, \bibinfo {author}
  {\bibfnamefont {K.~E.}\ \bibnamefont {Schmidt}},\ and\ \bibinfo {author}
  {\bibfnamefont {R.~B.}\ \bibnamefont {Wiringa}},\ }\bibfield  {title}
  {\bibinfo {title} {Quantum {M}onte {C}arlo methods for nuclear physics},\
  }\href {https://doi.org/10.1103/RevModPhys.87.1067} {\bibfield  {journal}
  {\bibinfo  {journal} {Rev. Mod. Phys.}\ }\textbf {\bibinfo {volume} {87}},\
  \bibinfo {pages} {1067} (\bibinfo {year} {2015})}\BibitemShut {NoStop}\bibitem [{\citenamefont {Odell}\ \emph {et~al.}(2016)\citenamefont {Odell},
  \citenamefont {Papenbrock},\ and\ \citenamefont
  {Platter}}]{odell2016:ir-extrap-quadrupole}\BibitemOpen
  \bibfield  {author} {\bibinfo {author} {\bibfnamefont {D.}~\bibnamefont
  {Odell}}, \bibinfo {author} {\bibfnamefont {T.}~\bibnamefont {Papenbrock}},\
  and\ \bibinfo {author} {\bibfnamefont {L.}~\bibnamefont {Platter}},\
  }\bibfield  {title} {\bibinfo {title} {Infrared extrapolations of quadrupole
  moments and transitions},\ }\href
  {https://doi.org/10.1103/PhysRevC.93.044331} {\bibfield  {journal} {\bibinfo
  {journal} {Phys. Rev. C}\ }\textbf {\bibinfo {volume} {93}},\ \bibinfo
  {pages} {044331} (\bibinfo {year} {2016})}\BibitemShut {NoStop}\bibitem [{\citenamefont {Barrett}\ \emph {et~al.}(2013)\citenamefont
  {Barrett}, \citenamefont {Navr\'{a}til},\ and\ \citenamefont
  {Vary}}]{barrett2013:ncsm}\BibitemOpen
  \bibfield  {author} {\bibinfo {author} {\bibfnamefont {B.~R.}\ \bibnamefont
  {Barrett}}, \bibinfo {author} {\bibfnamefont {P.}~\bibnamefont
  {Navr\'{a}til}},\ and\ \bibinfo {author} {\bibfnamefont {J.~P.}\ \bibnamefont
  {Vary}},\ }\bibfield  {title} {\bibinfo {title} {\textit{Ab initio} no core
  shell model},\ }\href {https://doi.org/10.1016/j.ppnp.2012.10.003} {\bibfield
   {journal} {\bibinfo  {journal} {Prog. Part. Nucl. Phys.}\ }\textbf {\bibinfo
  {volume} {69}},\ \bibinfo {pages} {131} (\bibinfo {year} {2013})}\BibitemShut
  {NoStop}\bibitem [{\citenamefont {Bogner}\ \emph {et~al.}(2008)\citenamefont {Bogner},
  \citenamefont {Furnstahl}, \citenamefont {Maris}, \citenamefont {Perry},
  \citenamefont {Schwenk},\ and\ \citenamefont
  {Vary}}]{bogner2008:ncsm-converg-2N}\BibitemOpen
  \bibfield  {author} {\bibinfo {author} {\bibfnamefont {S.~K.}\ \bibnamefont
  {Bogner}}, \bibinfo {author} {\bibfnamefont {R.~J.}\ \bibnamefont
  {Furnstahl}}, \bibinfo {author} {\bibfnamefont {P.}~\bibnamefont {Maris}},
  \bibinfo {author} {\bibfnamefont {R.~J.}\ \bibnamefont {Perry}}, \bibinfo
  {author} {\bibfnamefont {A.}~\bibnamefont {Schwenk}},\ and\ \bibinfo {author}
  {\bibfnamefont {J.}~\bibnamefont {Vary}},\ }\bibfield  {title} {\bibinfo
  {title} {Convergence in the no-core shell model with low-momentum two-nucleon
  interactions},\ }\href {https://doi.org/10.1016/j.nuclphysa.2007.12.008}
  {\bibfield  {journal} {\bibinfo  {journal} {Nucl. Phys. A}\ }\textbf
  {\bibinfo {volume} {801}},\ \bibinfo {pages} {21} (\bibinfo {year}
  {2008})}\BibitemShut {NoStop}\bibitem [{\citenamefont {Calci}\ and\ \citenamefont
  {Roth}(2016)}]{calci2016:observable-correlations-chiral}\BibitemOpen
  \bibfield  {author} {\bibinfo {author} {\bibfnamefont {A.}~\bibnamefont
  {Calci}}\ and\ \bibinfo {author} {\bibfnamefont {R.}~\bibnamefont {Roth}},\
  }\bibfield  {title} {\bibinfo {title} {Sensitivities and correlations of
  nuclear structure observables emerging from chiral interactions},\ }\href
  {https://doi.org/10.1103/PhysRevC.94.014322} {\bibfield  {journal} {\bibinfo
  {journal} {Phys. Rev. C}\ }\textbf {\bibinfo {volume} {94}},\ \bibinfo
  {pages} {014322} (\bibinfo {year} {2016})}\BibitemShut {NoStop}\bibitem [{\citenamefont {Caprio}\ \emph {et~al.}(2019)\citenamefont {Caprio},
  \citenamefont {Fasano}, \citenamefont {McCoy}, \citenamefont {Maris},\ and\
  \citenamefont {Vary}}]{caprio2019:bebands-sdanca19}\BibitemOpen
  \bibfield  {author} {\bibinfo {author} {\bibfnamefont {M.~A.}\ \bibnamefont
  {Caprio}}, \bibinfo {author} {\bibfnamefont {P.~J.}\ \bibnamefont {Fasano}},
  \bibinfo {author} {\bibfnamefont {A.~E.}\ \bibnamefont {McCoy}}, \bibinfo
  {author} {\bibfnamefont {P.}~\bibnamefont {Maris}},\ and\ \bibinfo {author}
  {\bibfnamefont {J.~P.}\ \bibnamefont {Vary}},\ }\bibfield  {title} {\bibinfo
  {title} {\textit{Ab initio} rotation in {$\isotope[10]{Be}$}},\ }\href
  {https://www.bjp-bg.com/paper.php?id=1208} {\bibfield  {journal} {\bibinfo
  {journal} {Bulg. J. Phys.}\ }\textbf {\bibinfo {volume} {46}},\ \bibinfo
  {pages} {445} (\bibinfo {year} {2019})}\BibitemShut {NoStop}\bibitem [{\citenamefont {Caprio}\ \emph {et~al.}(2022)\citenamefont {Caprio},
  \citenamefont {McCoy}, \citenamefont {Fasano},\ and\ \citenamefont
  {Dytrych}}]{caprio2022:10be-shape-sdanca21}\BibitemOpen
  \bibfield  {author} {\bibinfo {author} {\bibfnamefont {M.~A.}\ \bibnamefont
  {Caprio}}, \bibinfo {author} {\bibfnamefont {A.~E.}\ \bibnamefont {McCoy}},
  \bibinfo {author} {\bibfnamefont {P.~J.}\ \bibnamefont {Fasano}},\ and\
  \bibinfo {author} {\bibfnamefont {T.}~\bibnamefont {Dytrych}},\ }\bibfield
  {title} {\bibinfo {title} {Symmetry and shape coexistence in
  $\isotope[10]{Be}$},\ }\href {https://doi.org/10.55318/bgjp.2022.49.1.057}
  {\bibfield  {journal} {\bibinfo  {journal} {Bulg. J. Phys.}\ }\textbf
  {\bibinfo {volume} {49}},\ \bibinfo {pages} {57} (\bibinfo {year}
  {2022})}\BibitemShut {NoStop}\bibitem [{\citenamefont {Henderson}\ \emph {et~al.}(2019)\citenamefont
  {Henderson}, \citenamefont {Ahn}, \citenamefont {Caprio}, \citenamefont
  {Fasano}, \citenamefont {Simon}, \citenamefont {Tan}, \citenamefont
  {O'Malley}, \citenamefont {Allen}, \citenamefont {Bardayan}, \citenamefont
  {Blankstein}, \citenamefont {Frentz}, \citenamefont {Hall}, \citenamefont
  {Kolata}, \citenamefont {McCoy}, \citenamefont {Moylan}, \citenamefont
  {Reingold}, \citenamefont {Strauss},\ and\ \citenamefont
  {Torres-Isea}}]{henderson2019:7be-coulex}\BibitemOpen
  \bibfield  {author} {\bibinfo {author} {\bibfnamefont {S.~L.}\ \bibnamefont
  {Henderson}}, \bibinfo {author} {\bibfnamefont {T.}~\bibnamefont {Ahn}},
  \bibinfo {author} {\bibfnamefont {M.~A.}\ \bibnamefont {Caprio}}, \bibinfo
  {author} {\bibfnamefont {P.~J.}\ \bibnamefont {Fasano}}, \bibinfo {author}
  {\bibfnamefont {A.}~\bibnamefont {Simon}}, \bibinfo {author} {\bibfnamefont
  {W.}~\bibnamefont {Tan}}, \bibinfo {author} {\bibfnamefont {P.}~\bibnamefont
  {O'Malley}}, \bibinfo {author} {\bibfnamefont {J.}~\bibnamefont {Allen}},
  \bibinfo {author} {\bibfnamefont {D.~W.}\ \bibnamefont {Bardayan}}, \bibinfo
  {author} {\bibfnamefont {D.}~\bibnamefont {Blankstein}}, \bibinfo {author}
  {\bibfnamefont {B.}~\bibnamefont {Frentz}}, \bibinfo {author} {\bibfnamefont
  {M.~R.}\ \bibnamefont {Hall}}, \bibinfo {author} {\bibfnamefont {J.~J.}\
  \bibnamefont {Kolata}}, \bibinfo {author} {\bibfnamefont {A.~E.}\
  \bibnamefont {McCoy}}, \bibinfo {author} {\bibfnamefont {S.}~\bibnamefont
  {Moylan}}, \bibinfo {author} {\bibfnamefont {C.~S.}\ \bibnamefont
  {Reingold}}, \bibinfo {author} {\bibfnamefont {S.~Y.}\ \bibnamefont
  {Strauss}},\ and\ \bibinfo {author} {\bibfnamefont {R.~O.}\ \bibnamefont
  {Torres-Isea}},\ }\bibfield  {title} {\bibinfo {title} {First measurement of
  the {$B(E2; 3/2^- \rightarrow 1/2^-)$} transition strength in
  {$\isotope[7]{Be}$}: Testing \textit{ab initio} predictions for {$A=7$}
  nuclei},\ }\href {https://doi.org/10.1103/PhysRevC.99.064320} {\bibfield
  {journal} {\bibinfo  {journal} {Phys. Rev. C}\ }\textbf {\bibinfo {volume}
  {99}},\ \bibinfo {pages} {064320} (\bibinfo {year} {2019})}\BibitemShut
  {NoStop}\bibitem [{\citenamefont {Caprio}\ \emph {et~al.}(2021)\citenamefont {Caprio},
  \citenamefont {Fasano}, \citenamefont {Maris},\ and\ \citenamefont
  {McCoy}}]{caprio2021:emratio}\BibitemOpen
  \bibfield  {author} {\bibinfo {author} {\bibfnamefont {M.~A.}\ \bibnamefont
  {Caprio}}, \bibinfo {author} {\bibfnamefont {P.~J.}\ \bibnamefont {Fasano}},
  \bibinfo {author} {\bibfnamefont {P.}~\bibnamefont {Maris}},\ and\ \bibinfo
  {author} {\bibfnamefont {A.~E.}\ \bibnamefont {McCoy}},\ }\bibfield  {title}
  {\bibinfo {title} {Quadrupole moments and proton-neutron structure in
  $p$-shell mirror nuclei},\ }\href
  {https://doi.org/10.1103/PhysRevC.104.034319} {\bibfield  {journal} {\bibinfo
   {journal} {Phys. Rev. C}\ }\textbf {\bibinfo {volume} {104}},\ \bibinfo
  {pages} {034319} (\bibinfo {year} {2021})}\BibitemShut {NoStop}\bibitem [{\citenamefont {Caprio}\ and\ \citenamefont
  {Fasano}(2022)}]{caprio:8li-trans-PREPRINT}\BibitemOpen
  \bibfield  {author} {\bibinfo {author} {\bibfnamefont {M.~A.}\ \bibnamefont
  {Caprio}}\ and\ \bibinfo {author} {\bibfnamefont {P.~J.}\ \bibnamefont
  {Fasano}},\ }\href@noop {} {\bibinfo {title} {\textit{Ab initio} prediction
  of ${E2}$ strengths in $\isotope[8]{Li}$ and its neighbors by normalization
  to the measured quadrupole moment}} (\bibinfo {year} {2022}),\ \Eprint
  {https://arxiv.org/abs/2206.05628} {arXiv:2206.05628 [nucl-th]} \BibitemShut
  {NoStop}\bibitem [{\citenamefont {Stone}(2016)}]{stone2016:e2-moments}\BibitemOpen
  \bibfield  {author} {\bibinfo {author} {\bibfnamefont {N.~J.}\ \bibnamefont
  {Stone}},\ }\bibfield  {title} {\bibinfo {title} {Table of nuclear electric
  quadrupole moments},\ }\href {https://doi.org/10.1016/j.adt.2015.12.002}
  {\bibfield  {journal} {\bibinfo  {journal} {At. Data Nucl. Data Tables}\
  }\textbf {\bibinfo {volume} {111-112}},\ \bibinfo {pages} {1} (\bibinfo
  {year} {2016})}\BibitemShut {NoStop}\bibitem [{\citenamefont {Sargsyan}\ \emph {et~al.}(2022)\citenamefont
  {Sargsyan}, \citenamefont {Launey}, \citenamefont {Burkey}, \citenamefont
  {Gallant}, \citenamefont {Scielzo}, \citenamefont {Savard}, \citenamefont
  {Mercenne}, \citenamefont {Dytrych}, \citenamefont {Langr}, \citenamefont
  {Varriano}, \citenamefont {Longfellow}, \citenamefont {Hirsh},\ and\
  \citenamefont {Draayer}}]{sargsyan2022:8li-clustering-beta-recoil}\BibitemOpen
  \bibfield  {author} {\bibinfo {author} {\bibfnamefont {G.~H.}\ \bibnamefont
  {Sargsyan}}, \bibinfo {author} {\bibfnamefont {K.~D.}\ \bibnamefont
  {Launey}}, \bibinfo {author} {\bibfnamefont {M.~T.}\ \bibnamefont {Burkey}},
  \bibinfo {author} {\bibfnamefont {A.~T.}\ \bibnamefont {Gallant}}, \bibinfo
  {author} {\bibfnamefont {N.~D.}\ \bibnamefont {Scielzo}}, \bibinfo {author}
  {\bibfnamefont {G.}~\bibnamefont {Savard}}, \bibinfo {author} {\bibfnamefont
  {A.}~\bibnamefont {Mercenne}}, \bibinfo {author} {\bibfnamefont
  {T.}~\bibnamefont {Dytrych}}, \bibinfo {author} {\bibfnamefont
  {D.}~\bibnamefont {Langr}}, \bibinfo {author} {\bibfnamefont
  {L.}~\bibnamefont {Varriano}}, \bibinfo {author} {\bibfnamefont
  {B.}~\bibnamefont {Longfellow}}, \bibinfo {author} {\bibfnamefont {T.~Y.}\
  \bibnamefont {Hirsh}},\ and\ \bibinfo {author} {\bibfnamefont {J.~P.}\
  \bibnamefont {Draayer}},\ }\bibfield  {title} {\bibinfo {title} {Impact of
  clustering on the $\isotope[8]{Li}$ $\beta$ decay and recoil form factors},\
  }\href {https://doi.org/10.1103/PhysRevLett.128.202503} {\bibfield  {journal}
  {\bibinfo  {journal} {Phys. Rev. Lett.}\ }\textbf {\bibinfo {volume} {128}},\
  \bibinfo {pages} {202503} (\bibinfo {year} {2022})}\BibitemShut {NoStop}\bibitem [{\citenamefont {Angeli}\ and\ \citenamefont
  {Marinova}(2013)}]{angeli2013:charge-radii}\BibitemOpen
  \bibfield  {author} {\bibinfo {author} {\bibfnamefont {I.}~\bibnamefont
  {Angeli}}\ and\ \bibinfo {author} {\bibfnamefont {K.~P.}\ \bibnamefont
  {Marinova}},\ }\bibfield  {title} {\bibinfo {title} {Table of experimental
  nuclear ground state charge radii: An update},\ }\href
  {https://doi.org/10.1016/j.adt.2011.12.006} {\bibfield  {journal} {\bibinfo
  {journal} {At. Data Nucl. Data Tables}\ }\textbf {\bibinfo {volume} {99}},\
  \bibinfo {pages} {69} (\bibinfo {year} {2013})}\BibitemShut {NoStop}\bibitem [{\citenamefont {Friar}\ \emph {et~al.}(1997)\citenamefont {Friar},
  \citenamefont {Martorell},\ and\ \citenamefont
  {Sprung}}]{friar1997:charge-radius-correction}\BibitemOpen
  \bibfield  {author} {\bibinfo {author} {\bibfnamefont {J.~L.}\ \bibnamefont
  {Friar}}, \bibinfo {author} {\bibfnamefont {J.}~\bibnamefont {Martorell}},\
  and\ \bibinfo {author} {\bibfnamefont {D.~W.~L.}\ \bibnamefont {Sprung}},\
  }\bibfield  {title} {\bibinfo {title} {Nuclear sizes and the isotope shift},\
  }\href {https://doi.org/10.1103/PhysRevA.56.4579} {\bibfield  {journal}
  {\bibinfo  {journal} {Phys. Rev. A}\ }\textbf {\bibinfo {volume} {56}},\
  \bibinfo {pages} {4579} (\bibinfo {year} {1997})}\BibitemShut {NoStop}\bibitem [{\citenamefont {Lu}\ \emph {et~al.}(2013)\citenamefont {Lu},
  \citenamefont {Mueller}, \citenamefont {Drake}, \citenamefont
  {N{\"o}rtersh{\"a}user}, \citenamefont {Pieper},\ and\ \citenamefont
  {Yan}}]{lu2013:laser-neutron-rich}\BibitemOpen
  \bibfield  {author} {\bibinfo {author} {\bibfnamefont {Z.-T.}\ \bibnamefont
  {Lu}}, \bibinfo {author} {\bibfnamefont {P.}~\bibnamefont {Mueller}},
  \bibinfo {author} {\bibfnamefont {G.~W.~F.}\ \bibnamefont {Drake}}, \bibinfo
  {author} {\bibfnamefont {W.}~\bibnamefont {N{\"o}rtersh{\"a}user}}, \bibinfo
  {author} {\bibfnamefont {S.~C.}\ \bibnamefont {Pieper}},\ and\ \bibinfo
  {author} {\bibfnamefont {Z.-C.}\ \bibnamefont {Yan}},\ }\bibfield  {title}
  {\bibinfo {title} {Laser probing of neutron-rich nuclei in light atoms},\
  }\href {https://doi.org/10.1103/RevModPhys.85.1383} {\bibfield  {journal}
  {\bibinfo  {journal} {Rev. Mod. Phys.}\ }\textbf {\bibinfo {volume} {85}},\
  \bibinfo {pages} {1383} (\bibinfo {year} {2013})}\BibitemShut {NoStop}\bibitem [{\citenamefont {Tilley}\ \emph {et~al.}(2002)\citenamefont {Tilley},
  \citenamefont {Cheves}, \citenamefont {Godwin}, \citenamefont {Hale},
  \citenamefont {Hofmann}, \citenamefont {Kelley}, \citenamefont {Sheu},\ and\
  \citenamefont {Weller}}]{npa2002:005-007}\BibitemOpen
  \bibfield  {author} {\bibinfo {author} {\bibfnamefont {D.~R.}\ \bibnamefont
  {Tilley}}, \bibinfo {author} {\bibfnamefont {C.~M.}\ \bibnamefont {Cheves}},
  \bibinfo {author} {\bibfnamefont {J.~L.}\ \bibnamefont {Godwin}}, \bibinfo
  {author} {\bibfnamefont {G.~M.}\ \bibnamefont {Hale}}, \bibinfo {author}
  {\bibfnamefont {H.~M.}\ \bibnamefont {Hofmann}}, \bibinfo {author}
  {\bibfnamefont {J.~H.}\ \bibnamefont {Kelley}}, \bibinfo {author}
  {\bibfnamefont {C.~G.}\ \bibnamefont {Sheu}},\ and\ \bibinfo {author}
  {\bibfnamefont {H.~R.}\ \bibnamefont {Weller}},\ }\bibfield  {title}
  {\bibinfo {title} {Energy levels of light nuclei {$A = 5, 6, 7$}},\ }\href
  {https://doi.org/10.1016/S0375-9474(02)00597-3} {\bibfield  {journal}
  {\bibinfo  {journal} {Nucl. Phys. A}\ }\textbf {\bibinfo {volume} {708}},\
  \bibinfo {pages} {3} (\bibinfo {year} {2002})}\BibitemShut {NoStop}\bibitem [{\citenamefont {Pastore}\ \emph {et~al.}(2013)\citenamefont
  {Pastore}, \citenamefont {Pieper}, \citenamefont {Schiavilla},\ and\
  \citenamefont {Wiringa}}]{pastore2013:qmc-em-alt9}\BibitemOpen
  \bibfield  {author} {\bibinfo {author} {\bibfnamefont {S.}~\bibnamefont
  {Pastore}}, \bibinfo {author} {\bibfnamefont {S.~C.}\ \bibnamefont {Pieper}},
  \bibinfo {author} {\bibfnamefont {R.}~\bibnamefont {Schiavilla}},\ and\
  \bibinfo {author} {\bibfnamefont {R.~B.}\ \bibnamefont {Wiringa}},\
  }\bibfield  {title} {\bibinfo {title} {Quantum {M}onte {C}arlo calculations
  of electromagnetic moments and transitions in {$A\le9$} nuclei with
  meson-exchange currents derived from chiral effective field theory},\ }\href
  {https://doi.org/10.1103/PhysRevC.87.035503} {\bibfield  {journal} {\bibinfo
  {journal} {Phys. Rev. C}\ }\textbf {\bibinfo {volume} {87}},\ \bibinfo
  {pages} {035503} (\bibinfo {year} {2013})}\BibitemShut {NoStop}\bibitem [{\citenamefont {McCutchan}\ \emph {et~al.}(2009)\citenamefont
  {McCutchan}, \citenamefont {Lister}, \citenamefont {Wiringa}, \citenamefont
  {Pieper}, \citenamefont {Seweryniak}, \citenamefont {Greene}, \citenamefont
  {Carpenter}, \citenamefont {Chiara}, \citenamefont {Janssens}, \citenamefont
  {Khoo}, \citenamefont {Lauritsen}, \citenamefont {Stefanescu},\ and\
  \citenamefont {Zhu}}]{mccutchan2009:10be-dsam-gfmc}\BibitemOpen
  \bibfield  {author} {\bibinfo {author} {\bibfnamefont {E.~A.}\ \bibnamefont
  {McCutchan}}, \bibinfo {author} {\bibfnamefont {C.~J.}\ \bibnamefont
  {Lister}}, \bibinfo {author} {\bibfnamefont {R.~B.}\ \bibnamefont {Wiringa}},
  \bibinfo {author} {\bibfnamefont {S.~C.}\ \bibnamefont {Pieper}}, \bibinfo
  {author} {\bibfnamefont {D.}~\bibnamefont {Seweryniak}}, \bibinfo {author}
  {\bibfnamefont {J.~P.}\ \bibnamefont {Greene}}, \bibinfo {author}
  {\bibfnamefont {M.~P.}\ \bibnamefont {Carpenter}}, \bibinfo {author}
  {\bibfnamefont {C.~J.}\ \bibnamefont {Chiara}}, \bibinfo {author}
  {\bibfnamefont {R.~V.~F.}\ \bibnamefont {Janssens}}, \bibinfo {author}
  {\bibfnamefont {T.~L.}\ \bibnamefont {Khoo}}, \bibinfo {author}
  {\bibfnamefont {T.}~\bibnamefont {Lauritsen}}, \bibinfo {author}
  {\bibfnamefont {I.}~\bibnamefont {Stefanescu}},\ and\ \bibinfo {author}
  {\bibfnamefont {S.}~\bibnamefont {Zhu}},\ }\bibfield  {title} {\bibinfo
  {title} {Precise electromagnetic tests of \textit{ab initio} calculations of
  light nuclei: States in $\isotope[10]{Be}$},\ }\href
  {https://doi.org/10.1103/PhysRevLett.103.192501} {\bibfield  {journal}
  {\bibinfo  {journal} {Phys. Rev. Lett.}\ }\textbf {\bibinfo {volume} {103}},\
  \bibinfo {pages} {192501} (\bibinfo {year} {2009})}\BibitemShut {NoStop}\bibitem [{\citenamefont {McCutchan}\ \emph {et~al.}(2012)\citenamefont
  {McCutchan}, \citenamefont {Lister}, \citenamefont {Pieper}, \citenamefont
  {Wiringa}, \citenamefont {Seweryniak}, \citenamefont {Greene}, \citenamefont
  {Bertone}, \citenamefont {Carpenter}, \citenamefont {Chiara}, \citenamefont
  {G\"urdal}, \citenamefont {Hoffman}, \citenamefont {Janssens}, \citenamefont
  {Khoo}, \citenamefont {Lauritsen},\ and\ \citenamefont
  {Zhu}}]{mccutchan2012:10c-dsam}\BibitemOpen
  \bibfield  {author} {\bibinfo {author} {\bibfnamefont {E.~A.}\ \bibnamefont
  {McCutchan}}, \bibinfo {author} {\bibfnamefont {C.~J.}\ \bibnamefont
  {Lister}}, \bibinfo {author} {\bibfnamefont {S.~C.}\ \bibnamefont {Pieper}},
  \bibinfo {author} {\bibfnamefont {R.~B.}\ \bibnamefont {Wiringa}}, \bibinfo
  {author} {\bibfnamefont {D.}~\bibnamefont {Seweryniak}}, \bibinfo {author}
  {\bibfnamefont {J.~P.}\ \bibnamefont {Greene}}, \bibinfo {author}
  {\bibfnamefont {P.~F.}\ \bibnamefont {Bertone}}, \bibinfo {author}
  {\bibfnamefont {M.~P.}\ \bibnamefont {Carpenter}}, \bibinfo {author}
  {\bibfnamefont {C.~J.}\ \bibnamefont {Chiara}}, \bibinfo {author}
  {\bibfnamefont {G.}~\bibnamefont {G\"urdal}}, \bibinfo {author}
  {\bibfnamefont {C.~R.}\ \bibnamefont {Hoffman}}, \bibinfo {author}
  {\bibfnamefont {R.~V.~F.}\ \bibnamefont {Janssens}}, \bibinfo {author}
  {\bibfnamefont {T.~L.}\ \bibnamefont {Khoo}}, \bibinfo {author}
  {\bibfnamefont {T.}~\bibnamefont {Lauritsen}},\ and\ \bibinfo {author}
  {\bibfnamefont {S.}~\bibnamefont {Zhu}},\ }\bibfield  {title} {\bibinfo
  {title} {Lifetime of the ${2}_{1}^{+}$ state in $\isotope[10]{C}$},\ }\href
  {https://doi.org/10.1103/PhysRevC.86.014312} {\bibfield  {journal} {\bibinfo
  {journal} {Phys. Rev. C}\ }\textbf {\bibinfo {volume} {86}},\ \bibinfo
  {pages} {014312} (\bibinfo {year} {2012})}\BibitemShut {NoStop}\bibitem [{\citenamefont {Weller}\ \emph {et~al.}(1985)\citenamefont {Weller},
  \citenamefont {Egelhof}, \citenamefont {\v{C}aplar}, \citenamefont {Karban},
  \citenamefont {Kr\"amer}, \citenamefont {M\"obius}, \citenamefont {Moroz},
  \citenamefont {Rusek}, \citenamefont {Steffens}, \citenamefont {Tungate},
  \citenamefont {Blatt}, \citenamefont {Koenig},\ and\ \citenamefont
  {Fick}}]{weller1985:7li-coulex}\BibitemOpen
  \bibfield  {author} {\bibinfo {author} {\bibfnamefont {A.}~\bibnamefont
  {Weller}}, \bibinfo {author} {\bibfnamefont {P.}~\bibnamefont {Egelhof}},
  \bibinfo {author} {\bibfnamefont {R.}~\bibnamefont {\v{C}aplar}}, \bibinfo
  {author} {\bibfnamefont {O.}~\bibnamefont {Karban}}, \bibinfo {author}
  {\bibfnamefont {D.}~\bibnamefont {Kr\"amer}}, \bibinfo {author}
  {\bibfnamefont {K.-H.}\ \bibnamefont {M\"obius}}, \bibinfo {author}
  {\bibfnamefont {Z.}~\bibnamefont {Moroz}}, \bibinfo {author} {\bibfnamefont
  {K.}~\bibnamefont {Rusek}}, \bibinfo {author} {\bibfnamefont
  {E.}~\bibnamefont {Steffens}}, \bibinfo {author} {\bibfnamefont
  {G.}~\bibnamefont {Tungate}}, \bibinfo {author} {\bibfnamefont
  {K.}~\bibnamefont {Blatt}}, \bibinfo {author} {\bibfnamefont
  {I.}~\bibnamefont {Koenig}},\ and\ \bibinfo {author} {\bibfnamefont
  {D.}~\bibnamefont {Fick}},\ }\bibfield  {title} {\bibinfo {title}
  {Electromagnetic excitation of aligned $\isotope[7]{Li}$ nuclei},\ }\href
  {https://doi.org/10.1103/PhysRevLett.55.480} {\bibfield  {journal} {\bibinfo
  {journal} {Phys. Rev. Lett.}\ }\textbf {\bibinfo {volume} {55}},\ \bibinfo
  {pages} {480} (\bibinfo {year} {1985})}\BibitemShut {NoStop}\bibitem [{\citenamefont {Shirokov}\ \emph {et~al.}(2016)\citenamefont
  {Shirokov}, \citenamefont {Shin}, \citenamefont {Kim}, \citenamefont
  {Sosonkina}, \citenamefont {Maris},\ and\ \citenamefont
  {Vary}}]{shirokov2016:nn-daejeon16}\BibitemOpen
  \bibfield  {author} {\bibinfo {author} {\bibfnamefont {A.~M.}\ \bibnamefont
  {Shirokov}}, \bibinfo {author} {\bibfnamefont {I.~J.}\ \bibnamefont {Shin}},
  \bibinfo {author} {\bibfnamefont {Y.}~\bibnamefont {Kim}}, \bibinfo {author}
  {\bibfnamefont {M.}~\bibnamefont {Sosonkina}}, \bibinfo {author}
  {\bibfnamefont {P.}~\bibnamefont {Maris}},\ and\ \bibinfo {author}
  {\bibfnamefont {J.~P.}\ \bibnamefont {Vary}},\ }\bibfield  {title} {\bibinfo
  {title} {{N3LO} {$NN$} interaction adjusted to light nuclei in \textit{ab
  exitu} approach},\ }\href {https://doi.org/10.1016/j.physletb.2016.08.006}
  {\bibfield  {journal} {\bibinfo  {journal} {Phys. Lett. B}\ }\textbf
  {\bibinfo {volume} {761}},\ \bibinfo {pages} {87} (\bibinfo {year}
  {2016})}\BibitemShut {NoStop}\bibitem [{\citenamefont {Aktulga}\ \emph {et~al.}(2014)\citenamefont
  {Aktulga}, \citenamefont {Yang}, \citenamefont {Ng}, \citenamefont {Maris},\
  and\ \citenamefont {Vary}}]{aktulga2014:mfdn-scalability}\BibitemOpen
  \bibfield  {author} {\bibinfo {author} {\bibfnamefont {H.~M.}\ \bibnamefont
  {Aktulga}}, \bibinfo {author} {\bibfnamefont {C.}~\bibnamefont {Yang}},
  \bibinfo {author} {\bibfnamefont {E.~G.}\ \bibnamefont {Ng}}, \bibinfo
  {author} {\bibfnamefont {P.}~\bibnamefont {Maris}},\ and\ \bibinfo {author}
  {\bibfnamefont {J.~P.}\ \bibnamefont {Vary}},\ }\bibfield  {title} {\bibinfo
  {title} {Improving the scalability of symmetric iterative eigensolver for
  multi-core platforms},\ }\href {https://doi.org/10.1002/cpe.3129} {\bibfield
  {journal} {\bibinfo  {journal} {Concurrency Computat.: Pract. Exper.}\
  }\textbf {\bibinfo {volume} {26}},\ \bibinfo {pages} {2631} (\bibinfo {year}
  {2014})}\BibitemShut {NoStop}\bibitem [{\citenamefont {Shao}\ \emph {et~al.}(2018)\citenamefont {Shao},
  \citenamefont {Aktulga}, \citenamefont {Yang}, \citenamefont {Ng},
  \citenamefont {Maris},\ and\ \citenamefont
  {Vary}}]{shao2018:ncci-preconditioned}\BibitemOpen
  \bibfield  {author} {\bibinfo {author} {\bibfnamefont {M.}~\bibnamefont
  {Shao}}, \bibinfo {author} {\bibfnamefont {H.~M.}\ \bibnamefont {Aktulga}},
  \bibinfo {author} {\bibfnamefont {C.}~\bibnamefont {Yang}}, \bibinfo {author}
  {\bibfnamefont {E.~G.}\ \bibnamefont {Ng}}, \bibinfo {author} {\bibfnamefont
  {P.}~\bibnamefont {Maris}},\ and\ \bibinfo {author} {\bibfnamefont {J.~P.}\
  \bibnamefont {Vary}},\ }\bibfield  {title} {\bibinfo {title} {Accelerating
  nuclear configuration interaction calculations through a preconditioned block
  iterative eigensolver},\ }\href {https://doi.org/10.1016/j.cpc.2017.09.004}
  {\bibfield  {journal} {\bibinfo  {journal} {Comput. Phys. Commun.}\ }\textbf
  {\bibinfo {volume} {222}},\ \bibinfo {pages} {1} (\bibinfo {year}
  {2018})}\BibitemShut {NoStop}\bibitem [{\citenamefont {Suhonen}(2007)}]{suhonen2007:nucleons-nucleus}\BibitemOpen
  \bibfield  {author} {\bibinfo {author} {\bibfnamefont {J.}~\bibnamefont
  {Suhonen}},\ }\href {https://doi.org/10.1007/978-3-540-48861-3} {\emph
  {\bibinfo {title} {From Nucleons to Nucleus}}}\ (\bibinfo  {publisher}
  {Springer-Verlag},\ \bibinfo {address} {Berlin},\ \bibinfo {year}
  {2007})\BibitemShut {NoStop}\bibitem [{\citenamefont {Wiringa}\ \emph {et~al.}(1995)\citenamefont
  {Wiringa}, \citenamefont {Stoks},\ and\ \citenamefont
  {Schiavilla}}]{wiringa1995:nn-av18}\BibitemOpen
  \bibfield  {author} {\bibinfo {author} {\bibfnamefont {R.~B.}\ \bibnamefont
  {Wiringa}}, \bibinfo {author} {\bibfnamefont {V.~G.~J.}\ \bibnamefont
  {Stoks}},\ and\ \bibinfo {author} {\bibfnamefont {R.}~\bibnamefont
  {Schiavilla}},\ }\bibfield  {title} {\bibinfo {title} {Accurate
  nucleon-nucleon potential with charge-independence breaking},\ }\href
  {https://doi.org/10.1103/PhysRevC.51.38} {\bibfield  {journal} {\bibinfo
  {journal} {Phys. Rev. C}\ }\textbf {\bibinfo {volume} {51}},\ \bibinfo
  {pages} {38} (\bibinfo {year} {1995})}\BibitemShut {NoStop}\bibitem [{\citenamefont {Pieper}(2008)}]{pieper2008:3n-il7-fm50}\BibitemOpen
  \bibfield  {author} {\bibinfo {author} {\bibfnamefont {S.~C.}\ \bibnamefont
  {Pieper}},\ }\bibfield  {title} {\bibinfo {title} {The {I}llinois extension
  to the {F}ujita-{M}iyazawa three-nucleon force},\ }in\ \href
  {https://doi.org/10.1063/1.2932280} {\emph {\bibinfo {booktitle} {New Facet
  of Three Nucleon Force --- 50 Years of Fujita Miyazawa Three Nucleon Force
  {(FM50)}: Proceedings of the International Symposium on New Facet of Three
  Nucleon Force}}},\ \bibinfo {series and number} {\bibinfo {series} {AIP Conf.
  Proc.}\ No.\ \bibinfo {number} {1011}},\ \bibinfo {editor} {edited by\
  \bibinfo {editor} {\bibfnamefont {H.}~\bibnamefont {Sakai}}, \bibinfo
  {editor} {\bibfnamefont {K.}~\bibnamefont {Sekiguchi}},\ and\ \bibinfo
  {editor} {\bibfnamefont {B.~F.}\ \bibnamefont {Gibson}}}\ (\bibinfo
  {publisher} {AIP},\ \bibinfo {address} {New York},\ \bibinfo {year} {2008})\
  pp.\ \bibinfo {pages} {143--152}\BibitemShut {NoStop}\bibitem [{\citenamefont {Kanada-En'yo}\ \emph {et~al.}(1999)\citenamefont
  {Kanada-En'yo}, \citenamefont {Horiuchi},\ and\ \citenamefont
  {Dot{\'e}}}]{kanadaenyo1999:10be-amd}\BibitemOpen
  \bibfield  {author} {\bibinfo {author} {\bibfnamefont {Y.}~\bibnamefont
  {Kanada-En'yo}}, \bibinfo {author} {\bibfnamefont {H.}~\bibnamefont
  {Horiuchi}},\ and\ \bibinfo {author} {\bibfnamefont {A.}~\bibnamefont
  {Dot{\'e}}},\ }\bibfield  {title} {\bibinfo {title} {Structure of excited
  states of $\isotope[10]{Be}$ studied with antisymmetrized molecular
  dynamics},\ }\href {https://doi.org/10.1103/PhysRevC.60.064304} {\bibfield
  {journal} {\bibinfo  {journal} {Phys. Rev. C}\ }\textbf {\bibinfo {volume}
  {60}},\ \bibinfo {pages} {064304} (\bibinfo {year} {1999})}\BibitemShut
  {NoStop}\bibitem [{\citenamefont {Bohlen}\ \emph {et~al.}(2007)\citenamefont {Bohlen},
  \citenamefont {Dorsch}, \citenamefont {Kokalova}, \citenamefont {von
  Oertzen}, \citenamefont {Schulz},\ and\ \citenamefont
  {Wheldon}}]{bohlen2007:10be-pickup}\BibitemOpen
  \bibfield  {author} {\bibinfo {author} {\bibfnamefont {H.~G.}\ \bibnamefont
  {Bohlen}}, \bibinfo {author} {\bibfnamefont {T.}~\bibnamefont {Dorsch}},
  \bibinfo {author} {\bibfnamefont {{\mbox{Tz}}.}~\bibnamefont {Kokalova}},
  \bibinfo {author} {\bibfnamefont {W.}~\bibnamefont {von Oertzen}}, \bibinfo
  {author} {\bibfnamefont {{\mbox{Ch}}.}~\bibnamefont {Schulz}},\ and\ \bibinfo
  {author} {\bibfnamefont {C.}~\bibnamefont {Wheldon}},\ }\bibfield  {title}
  {\bibinfo {title} {Structure of $\isotope[10]{Be}$ from the
  $\isotope[12]{C}(\isotope[12]{C},\isotope[14]{O})\isotope[10]{Be}$
  reaction},\ }\href {https://doi.org/10.1103/PhysRevC.75.054604} {\bibfield
  {journal} {\bibinfo  {journal} {Phys. Rev. C}\ }\textbf {\bibinfo {volume}
  {75}},\ \bibinfo {pages} {054604} (\bibinfo {year} {2007})}\BibitemShut
  {NoStop}\bibitem [{\citenamefont {Kanada-En'yo}\ and\ \citenamefont
  {Horiuchi}(1997)}]{kanadaenyo1997:c-amd-pn-decoupling}\BibitemOpen
  \bibfield  {author} {\bibinfo {author} {\bibfnamefont {Y.}~\bibnamefont
  {Kanada-En'yo}}\ and\ \bibinfo {author} {\bibfnamefont {H.}~\bibnamefont
  {Horiuchi}},\ }\bibfield  {title} {\bibinfo {title} {Opposite deformations
  between protons and neutrons in proton-rich $\isotope{C}$ isotopes},\ }\href
  {https://doi.org/10.1103/PhysRevC.55.2860} {\bibfield  {journal} {\bibinfo
  {journal} {Phys. Rev. C}\ }\textbf {\bibinfo {volume} {55}},\ \bibinfo
  {pages} {2860} (\bibinfo {year} {1997})}\BibitemShut {NoStop}\bibitem [{\citenamefont {Suhara}\ and\ \citenamefont
  {Kanada-En'yo}(2010)}]{suhara2010:amd-deformation}\BibitemOpen
  \bibfield  {author} {\bibinfo {author} {\bibfnamefont {T.}~\bibnamefont
  {Suhara}}\ and\ \bibinfo {author} {\bibfnamefont {Y.}~\bibnamefont
  {Kanada-En'yo}},\ }\bibfield  {title} {\bibinfo {title} {Quadrupole
  deformation $\beta$ and $\gamma$ constraint in a framework of antisymmetrized
  molecular dynamics},\ }\href {https://doi.org/10.1143/PTP.123.303} {\bibfield
   {journal} {\bibinfo  {journal} {Prog. Theor. Phys.}\ }\textbf {\bibinfo
  {volume} {123}},\ \bibinfo {pages} {303} (\bibinfo {year}
  {2010})}\BibitemShut {NoStop}\bibitem [{\citenamefont {Davydov}\ and\ \citenamefont
  {Filippov}(1958)}]{davydov1958:arm-intro}\BibitemOpen
  \bibfield  {author} {\bibinfo {author} {\bibfnamefont {A.~S.}\ \bibnamefont
  {Davydov}}\ and\ \bibinfo {author} {\bibfnamefont {G.~F.}\ \bibnamefont
  {Filippov}},\ }\bibfield  {title} {\bibinfo {title} {Rotational states in
  even atomic nuclei},\ }\href {https://doi.org/10.1016/0029-5582(58)90153-6}
  {\bibfield  {journal} {\bibinfo  {journal} {Nucl. Phys.}\ }\textbf {\bibinfo
  {volume} {8}},\ \bibinfo {pages} {237} (\bibinfo {year} {1958})}\BibitemShut
  {NoStop}\bibitem [{\citenamefont
  {{Meyer-ter-Vehn}}(1975)}]{meyertervehn1975:triax-odda}\BibitemOpen
  \bibfield  {author} {\bibinfo {author} {\bibfnamefont {J.}~\bibnamefont
  {{Meyer-ter-Vehn}}},\ }\bibfield  {title} {\bibinfo {title} {Collective model
  description of transitional odd-{$A$} nuclei: {(I)}. {T}he
  triaxial-rotor-plus-particle model},\ }\href
  {https://doi.org/10.1016/0375-9474(75)90095-0} {\bibfield  {journal}
  {\bibinfo  {journal} {Nucl. Phys. A}\ }\textbf {\bibinfo {volume} {249}},\
  \bibinfo {pages} {111} (\bibinfo {year} {1975})}\BibitemShut {NoStop}\bibitem [{\citenamefont {Warburton}\ \emph {et~al.}(1966)\citenamefont
  {Warburton}, \citenamefont {Olness}, \citenamefont {Jones}, \citenamefont
  {Chasman}, \citenamefont {Ristinen},\ and\ \citenamefont
  {Wilkinson}}]{warburton1966:a10-a11-a12-dsam}\BibitemOpen
  \bibfield  {author} {\bibinfo {author} {\bibfnamefont {E.~K.}\ \bibnamefont
  {Warburton}}, \bibinfo {author} {\bibfnamefont {J.~W.}\ \bibnamefont
  {Olness}}, \bibinfo {author} {\bibfnamefont {K.~W.}\ \bibnamefont {Jones}},
  \bibinfo {author} {\bibfnamefont {C.}~\bibnamefont {Chasman}}, \bibinfo
  {author} {\bibfnamefont {R.~A.}\ \bibnamefont {Ristinen}},\ and\ \bibinfo
  {author} {\bibfnamefont {D.~H.}\ \bibnamefont {Wilkinson}},\ }\bibfield
  {title} {\bibinfo {title} {Lifetime determinations for nuclei {$A = 10$},
  $11$, and $12$ from gamma-ray {D}oppler shifts},\ }\href
  {https://doi.org/10.1103/PhysRev.148.1072} {\bibfield  {journal} {\bibinfo
  {journal} {Phys. Rev.}\ }\textbf {\bibinfo {volume} {148}},\ \bibinfo {pages}
  {1072} (\bibinfo {year} {1966})}\BibitemShut {NoStop}\bibitem [{\citenamefont {Fisher}\ \emph {et~al.}(1968)\citenamefont {Fisher},
  \citenamefont {Hanna}, \citenamefont {Healey},\ and\ \citenamefont
  {Paul}}]{fisher1968:a10-dsam}\BibitemOpen
  \bibfield  {author} {\bibinfo {author} {\bibfnamefont {T.~R.}\ \bibnamefont
  {Fisher}}, \bibinfo {author} {\bibfnamefont {S.~S.}\ \bibnamefont {Hanna}},
  \bibinfo {author} {\bibfnamefont {D.~C.}\ \bibnamefont {Healey}},\ and\
  \bibinfo {author} {\bibfnamefont {P.}~\bibnamefont {Paul}},\ }\bibfield
  {title} {\bibinfo {title} {Lifetimes of levels in {$A = 10$} nuclei},\ }\href
  {https://doi.org/10.1103/PhysRev.176.1130} {\bibfield  {journal} {\bibinfo
  {journal} {Phys. Rev.}\ }\textbf {\bibinfo {volume} {176}},\ \bibinfo {pages}
  {1130} (\bibinfo {year} {1968})}\BibitemShut {NoStop}\bibitem [{\citenamefont {Pieper}\ and\ \citenamefont
  {Carlson}()}]{pieper:cited}\BibitemOpen
  \bibfield  {author} {\bibinfo {author} {\bibfnamefont {S.~C.}\ \bibnamefont
  {Pieper}}\ and\ \bibinfo {author} {\bibfnamefont {J.}~\bibnamefont
  {Carlson}},\ }\href@noop {} {}\bibinfo {note} {{a}s cited in
  Ref.~\cite{carlson2015:qmc-nuclear}}\BibitemShut {NoStop}\bibitem [{\citenamefont {Shirokov}\ \emph {et~al.}(2007)\citenamefont
  {Shirokov}, \citenamefont {Vary}, \citenamefont {Mazur},\ and\ \citenamefont
  {Weber}}]{shirokov2007:nn-jisp16}\BibitemOpen
  \bibfield  {author} {\bibinfo {author} {\bibfnamefont {A.~M.}\ \bibnamefont
  {Shirokov}}, \bibinfo {author} {\bibfnamefont {J.~P.}\ \bibnamefont {Vary}},
  \bibinfo {author} {\bibfnamefont {A.~I.}\ \bibnamefont {Mazur}},\ and\
  \bibinfo {author} {\bibfnamefont {T.~A.}\ \bibnamefont {Weber}},\ }\bibfield
  {title} {\bibinfo {title} {Realistic nuclear {H}amiltonian: \textit{Ab exitu}
  approach},\ }\href {https://doi.org/10.1016/j.physletb.2006.10.066}
  {\bibfield  {journal} {\bibinfo  {journal} {Phys. Lett. B}\ }\textbf
  {\bibinfo {volume} {644}},\ \bibinfo {pages} {33} (\bibinfo {year}
  {2007})}\BibitemShut {NoStop}\bibitem [{\citenamefont {Epelbaum}\ \emph
  {et~al.}(2015{\natexlab{a}})\citenamefont {Epelbaum}, \citenamefont {Krebs},\
  and\ \citenamefont {Mei\ss{}ner}}]{epelbaum2015:lenpic-n4lo-scs}\BibitemOpen
  \bibfield  {author} {\bibinfo {author} {\bibfnamefont {E.}~\bibnamefont
  {Epelbaum}}, \bibinfo {author} {\bibfnamefont {H.}~\bibnamefont {Krebs}},\
  and\ \bibinfo {author} {\bibfnamefont {U.-G.}\ \bibnamefont {Mei\ss{}ner}},\
  }\bibfield  {title} {\bibinfo {title} {Precision nucleon-nucleon potential at
  fifth order in the chiral expansion},\ }\href
  {https://doi.org/10.1103/PhysRevLett.115.122301} {\bibfield  {journal}
  {\bibinfo  {journal} {Phys. Rev. Lett.}\ }\textbf {\bibinfo {volume} {115}},\
  \bibinfo {pages} {122301} (\bibinfo {year} {2015}{\natexlab{a}})}\BibitemShut
  {NoStop}\bibitem [{\citenamefont {Epelbaum}\ \emph
  {et~al.}(2015{\natexlab{b}})\citenamefont {Epelbaum}, \citenamefont {Krebs},\
  and\ \citenamefont {Mei\ss{}ner}}]{epelbaum2015:lenpic-n3lo-scs}\BibitemOpen
  \bibfield  {author} {\bibinfo {author} {\bibfnamefont {E.}~\bibnamefont
  {Epelbaum}}, \bibinfo {author} {\bibfnamefont {H.}~\bibnamefont {Krebs}},\
  and\ \bibinfo {author} {\bibfnamefont {U.-G.}\ \bibnamefont {Mei\ss{}ner}},\
  }\bibfield  {title} {\bibinfo {title} {Improved chiral nucleon-nucleon
  potential up to next-to-next-to-next-to-leading order},\ }\href
  {https://doi.org/10.1140/epja/i2015-15053-8} {\bibfield  {journal} {\bibinfo
  {journal} {Eur. Phys. J. A}\ }\textbf {\bibinfo {volume} {51}},\ \bibinfo
  {pages} {53} (\bibinfo {year} {2015}{\natexlab{b}})}\BibitemShut {NoStop}\bibitem [{\citenamefont {Tanihata}\ \emph {et~al.}(1985)\citenamefont
  {Tanihata}, \citenamefont {Hamagaki}, \citenamefont {Hashimoto},
  \citenamefont {Shida}, \citenamefont {Yoshikawa}, \citenamefont {Sugimoto},
  \citenamefont {Yamakawa}, \citenamefont {Kobayashi},\ and\ \citenamefont
  {Takahashi}}]{tanihata1985:radii-11li-halo}\BibitemOpen
  \bibfield  {author} {\bibinfo {author} {\bibfnamefont {I.}~\bibnamefont
  {Tanihata}}, \bibinfo {author} {\bibfnamefont {H.}~\bibnamefont {Hamagaki}},
  \bibinfo {author} {\bibfnamefont {O.}~\bibnamefont {Hashimoto}}, \bibinfo
  {author} {\bibfnamefont {Y.}~\bibnamefont {Shida}}, \bibinfo {author}
  {\bibfnamefont {N.}~\bibnamefont {Yoshikawa}}, \bibinfo {author}
  {\bibfnamefont {K.}~\bibnamefont {Sugimoto}}, \bibinfo {author}
  {\bibfnamefont {O.}~\bibnamefont {Yamakawa}}, \bibinfo {author}
  {\bibfnamefont {T.}~\bibnamefont {Kobayashi}},\ and\ \bibinfo {author}
  {\bibfnamefont {N.}~\bibnamefont {Takahashi}},\ }\bibfield  {title} {\bibinfo
  {title} {Measurements of interaction cross sections and nuclear radii in the
  light $p$-shell region},\ }\href
  {https://doi.org/10.1103/PhysRevLett.55.2676} {\bibfield  {journal} {\bibinfo
   {journal} {Phys. Rev. Lett.}\ }\textbf {\bibinfo {volume} {55}},\ \bibinfo
  {pages} {2676} (\bibinfo {year} {1985})}\BibitemShut {NoStop}\bibitem [{\citenamefont {Jonson}(2004)}]{jonson2004:light-dripline}\BibitemOpen
  \bibfield  {author} {\bibinfo {author} {\bibfnamefont {B.}~\bibnamefont
  {Jonson}},\ }\bibfield  {title} {\bibinfo {title} {Light dripline nuclei},\
  }\href {https://doi.org/10.1016/j.physrep.2003.07.004} {\bibfield  {journal}
  {\bibinfo  {journal} {Phys. Rep.}\ }\textbf {\bibinfo {volume} {389}},\
  \bibinfo {pages} {1} (\bibinfo {year} {2004})}\BibitemShut {NoStop}\bibitem [{\citenamefont {Vermeer}\ \emph {et~al.}(1983)\citenamefont
  {Vermeer}, \citenamefont {Esat}, \citenamefont {Kuehner}, \citenamefont
  {Spear}, \citenamefont {Baxter},\ and\ \citenamefont
  {Hinds}}]{vermeer1983:12c-coulex}\BibitemOpen
  \bibfield  {author} {\bibinfo {author} {\bibfnamefont {W.~J.}\ \bibnamefont
  {Vermeer}}, \bibinfo {author} {\bibfnamefont {M.~T.}\ \bibnamefont {Esat}},
  \bibinfo {author} {\bibfnamefont {J.~A.}\ \bibnamefont {Kuehner}}, \bibinfo
  {author} {\bibfnamefont {R.~H.}\ \bibnamefont {Spear}}, \bibinfo {author}
  {\bibfnamefont {A.~M.}\ \bibnamefont {Baxter}},\ and\ \bibinfo {author}
  {\bibfnamefont {S.}~\bibnamefont {Hinds}},\ }\bibfield  {title} {\bibinfo
  {title} {Electric quadrupole moment of the first excited state of
  $\isotope[12]{C}$},\ }\href {https://doi.org/10.1016/0370-2693(83)91160-7}
  {\bibfield  {journal} {\bibinfo  {journal} {Phys. Lett. B}\ }\textbf
  {\bibinfo {volume} {122}},\ \bibinfo {pages} {23} (\bibinfo {year}
  {1983})}\BibitemShut {NoStop}\bibitem [{\citenamefont {Vorabbi}\ \emph {et~al.}(2019)\citenamefont
  {Vorabbi}, \citenamefont {Navr\'atil}, \citenamefont {Quaglioni},\ and\
  \citenamefont {Hupin}}]{vorabbi2019:7be-7li-ncsmc}\BibitemOpen
  \bibfield  {author} {\bibinfo {author} {\bibfnamefont {M.}~\bibnamefont
  {Vorabbi}}, \bibinfo {author} {\bibfnamefont {P.}~\bibnamefont {Navr\'atil}},
  \bibinfo {author} {\bibfnamefont {S.}~\bibnamefont {Quaglioni}},\ and\
  \bibinfo {author} {\bibfnamefont {G.}~\bibnamefont {Hupin}},\ }\bibfield
  {title} {\bibinfo {title} {$\isotope[7]{Be}$ and $\isotope[7]{Li}$ nuclei
  within the no-core shell model with continuum},\ }\href
  {https://doi.org/10.1103/PhysRevC.100.024304} {\bibfield  {journal} {\bibinfo
   {journal} {Phys. Rev. C}\ }\textbf {\bibinfo {volume} {100}},\ \bibinfo
  {pages} {024304} (\bibinfo {year} {2019})}\BibitemShut {NoStop}\bibitem [{\citenamefont {Rowe}(1985)}]{rowe1985:micro-collective-sp6r}\BibitemOpen
  \bibfield  {author} {\bibinfo {author} {\bibfnamefont {D.~J.}\ \bibnamefont
  {Rowe}},\ }\bibfield  {title} {\bibinfo {title} {Microscopic theory of the
  nuclear collective model},\ }\href
  {https://doi.org/10.1088/0034-4885/48/10/003} {\bibfield  {journal} {\bibinfo
   {journal} {Rep. Prog. Phys.}\ }\textbf {\bibinfo {volume} {48}},\ \bibinfo
  {pages} {1419} (\bibinfo {year} {1985})}\BibitemShut {NoStop}\bibitem [{\citenamefont {Dytrych}\ \emph {et~al.}(2008)\citenamefont
  {Dytrych}, \citenamefont {Sviratcheva}, \citenamefont {Draayer},
  \citenamefont {Bahri},\ and\ \citenamefont {Vary}}]{dytrych2008:sp-ncsm}\BibitemOpen
  \bibfield  {author} {\bibinfo {author} {\bibfnamefont {T.}~\bibnamefont
  {Dytrych}}, \bibinfo {author} {\bibfnamefont {K.~D.}\ \bibnamefont
  {Sviratcheva}}, \bibinfo {author} {\bibfnamefont {J.~P.}\ \bibnamefont
  {Draayer}}, \bibinfo {author} {\bibfnamefont {C.}~\bibnamefont {Bahri}},\
  and\ \bibinfo {author} {\bibfnamefont {J.~P.}\ \bibnamefont {Vary}},\
  }\bibfield  {title} {\bibinfo {title} {\textit{Ab initio} symplectic no-core
  shell model},\ }\href {https://doi.org/10.1088/0954-3899/35/12/123101}
  {\bibfield  {journal} {\bibinfo  {journal} {J. Phys. G}\ }\textbf {\bibinfo
  {volume} {35}},\ \bibinfo {pages} {123101} (\bibinfo {year}
  {2008})}\BibitemShut {NoStop}\bibitem [{\citenamefont {McCoy}\ \emph {et~al.}(2020)\citenamefont {McCoy},
  \citenamefont {Caprio}, \citenamefont {Dytrych},\ and\ \citenamefont
  {Fasano}}]{mccoy2020:spfamilies}\BibitemOpen
  \bibfield  {author} {\bibinfo {author} {\bibfnamefont {A.~E.}\ \bibnamefont
  {McCoy}}, \bibinfo {author} {\bibfnamefont {M.~A.}\ \bibnamefont {Caprio}},
  \bibinfo {author} {\bibfnamefont {T.}~\bibnamefont {Dytrych}},\ and\ \bibinfo
  {author} {\bibfnamefont {P.~J.}\ \bibnamefont {Fasano}},\ }\bibfield  {title}
  {\bibinfo {title} {Emergent {$\grpsptr$} dynamical symmetry in the nuclear
  many-body system from an \emph{ab initio} description},\ }\href
  {https://doi.org/10.1103/PhysRevLett.125.102505} {\bibfield  {journal}
  {\bibinfo  {journal} {Phys. Rev. Lett.}\ }\textbf {\bibinfo {volume} {125}},\
  \bibinfo {pages} {102505} (\bibinfo {year} {2020})}\BibitemShut {NoStop}\bibitem [{\citenamefont {Constantinou}\ \emph {et~al.}(2017)\citenamefont
  {Constantinou}, \citenamefont {Caprio}, \citenamefont {Vary},\ and\
  \citenamefont {Maris}}]{constantinou2017:natorb-natowitz16}\BibitemOpen
  \bibfield  {author} {\bibinfo {author} {\bibfnamefont
  {{\mbox{Ch}}.}~\bibnamefont {Constantinou}}, \bibinfo {author} {\bibfnamefont
  {M.~A.}\ \bibnamefont {Caprio}}, \bibinfo {author} {\bibfnamefont {J.~P.}\
  \bibnamefont {Vary}},\ and\ \bibinfo {author} {\bibfnamefont
  {P.}~\bibnamefont {Maris}},\ }\bibfield  {title} {\bibinfo {title}
  {\textit{Ab initio} properties of the halo nucleus $\isotope[6]{He}$ in a
  natural orbital basis},\ }\href {https://doi.org/10.1007/s41365-017-0332-6}
  {\bibfield  {journal} {\bibinfo  {journal} {Nucl. Sci. Techniques}\ }\textbf
  {\bibinfo {volume} {28}},\ \bibinfo {pages} {179} (\bibinfo {year}
  {2017})}\BibitemShut {NoStop}\bibitem [{\citenamefont {Fasano}\ \emph {et~al.}(2022)\citenamefont {Fasano},
  \citenamefont {Constantinou}, \citenamefont {Caprio}, \citenamefont {Maris},\
  and\ \citenamefont {Vary}}]{fasano2022:natorb}\BibitemOpen
  \bibfield  {author} {\bibinfo {author} {\bibfnamefont {P.~J.}\ \bibnamefont
  {Fasano}}, \bibinfo {author} {\bibfnamefont {{\mbox{Ch}}.}~\bibnamefont
  {Constantinou}}, \bibinfo {author} {\bibfnamefont {M.~A.}\ \bibnamefont
  {Caprio}}, \bibinfo {author} {\bibfnamefont {P.}~\bibnamefont {Maris}},\ and\
  \bibinfo {author} {\bibfnamefont {J.~P.}\ \bibnamefont {Vary}},\ }\bibfield
  {title} {\bibinfo {title} {Natural orbitals for the \textit{ab initio}
  no-core configuration interaction approach},\ }\href
  {https://doi.org/10.1103/PhysRevC.105.054301} {\bibfield  {journal} {\bibinfo
   {journal} {Phys. Rev. C}\ }\textbf {\bibinfo {volume} {105}},\ \bibinfo
  {pages} {054301} (\bibinfo {year} {2022})}\BibitemShut {NoStop}\bibitem [{\citenamefont {Roth}\ and\ \citenamefont
  {Navr\'atil}(2007)}]{roth2007:it-ncsm-40ca}\BibitemOpen
  \bibfield  {author} {\bibinfo {author} {\bibfnamefont {R.}~\bibnamefont
  {Roth}}\ and\ \bibinfo {author} {\bibfnamefont {P.}~\bibnamefont
  {Navr\'atil}},\ }\bibfield  {title} {\bibinfo {title} {\textit{Ab initio}
  study of $\isotope[40]{Ca}$ with an importance-truncated no-core shell
  model},\ }\href {https://doi.org/10.1103/PhysRevLett.99.092501} {\bibfield
  {journal} {\bibinfo  {journal} {Phys. Rev. Lett.}\ }\textbf {\bibinfo
  {volume} {99}},\ \bibinfo {pages} {092501} (\bibinfo {year}
  {2007})}\BibitemShut {NoStop}\end{thebibliography}
\nocite{control:title-on}

\end{document}